\documentclass[aps,prb,amsmath,amssymb,floatfix,twocolumn]{revtex4-1}
\usepackage{bbold}
\usepackage{graphicx}
\usepackage{subfigure}
\usepackage{color}
\usepackage{hyperref}
\usepackage{array}
\usepackage{times}

\begin{document}

\title{The magnetic phase diagram of an extended $J_1-J_2$ model on a modulated square lattice and its implications for the antiferromagnetic phase of $\mathrm{K}_{y}\mathrm{Fe}_{x}\mathrm{Se}_2$ }

\author{Rong Yu}
\affiliation{Department of Physics and Astronomy, Rice University, Houston,
TX 77005}
\author{Pallab Goswami}
\affiliation{Department of Physics and Astronomy, Rice University, Houston,
TX 77005}
\author{Qimiao Si}
\affiliation{Department of Physics and Astronomy, Rice University, Houston,
TX 77005}

\begin{abstract}
Motivated by the experimentally observed $\sqrt{5} \times \sqrt{5}$ iron vacancy order and a block spin antiferromagnetic phase with large magnetic moment in $\mathrm{K}_{0.8}\mathrm{Fe}_{1.6}\mathrm{Se}_2$, we study the magnetic phase diagram of an extended $J_1-J_2$ model on a $\frac{1}{5}$-depleted square lattice with $\sqrt{5} \times \sqrt{5}$ vacancy order, using a classical Monte Carlo analysis. The magnetic phase diagram involves various antiferromagnetically ordered phases, and most of them have higher order commensuration. We find that the experimentally relevant block-spin state occupies a significant portion of the phase diagram, and we discuss the spin dynamics of this phase using a linear spin-wave analysis.
By comparing the calculated magnetization with the experimental values of magnetic moment, we determine the physical parameter regimes corresponding to the block spin antiferromagnetic phase.
Based on our spin wave calculations in different parameter regimes, we show how spin-wave degeneracy along the high symmetry directions of the magnetic Brillouin zone can provide information regarding the underlying exchange couplings. We have also analyzed the magnetic phase diagram of a $J_1-J_2$ model on two different modulated square lattices relevant for $\mathrm{K}_y\mathrm{Fe}_{1.5}\mathrm{Se}_2$, which respectively
 exhibit $\frac{1}{4}$-depleted $2\times2$ and $4\times2$ vacancy ordering.
\end{abstract}

\maketitle


\section{Introduction}
Ever since the discovery of the high temperature superconductivity in iron pnictides in the vicinity of an antiferromagnetically ordered phase \cite{YKamihara, ZARen, Cruz}, the strength of electronic correlations in iron based superconductors has been a central issue of debate. Significant efforts based on both weak and strong coupling pictures have been made to explain the $(\pi, 0)$ magnetic order in the parent compounds, and the emergence
of superconductivity upon carrier doping. On one hand, the
$(\pi, 0)$ collinear antiferromagnetic order
arises in a weak coupling picture that relies on the presence of quasi-nested electron and hole pockets respectively at $M$ and $\Gamma$ points of the extended
Brillouin zone \cite{Dong}. On the other hand, the experimentally observed large electrical resistivity (``bad metal" with $k_F l \sim 1$), a strong suppression of Drude weight \cite{MQazilbash}, and the temperature-induced spectral weight transfer \cite{WZHu,AVBoris,JYang} have suggested sizable strength of electronic correlations. These experimental evidences have suggested the placement of the iron pnictides in close proximity of a putative Mott transition \cite{QSi1, QSi2, AKutepov}. In a metallic system close to a Mott transition, quasi-local moments are expected to arise, which for the iron pnictides are described in terms
of $J_1-J_2$ couplings \cite{QSi1,QSi2}; here $J_1$ and $J_2$ refer to nearest-neighbor and next-nearest-neighbor spin exchange interactions on the iron square lattice.
Hence in the strong coupling scenario, the $(\pi, 0)$ collinear
antiferromagnetic order in these materials can be described in terms of such a $J_1-J_2$ model.
This picture is further supported by the experimental observation of zone boundary spin wave excitations in the magnetically ordered state at low temperatures \cite{JZhao}. When protected by the broken symmetry, many low energy properties of the ordered state have been addressed using either strong or weak coupling approaches, and this has contributed to
 the dilemma over the strength of electronic correlations.

For 11-iron chalcogenides $\mathrm{Fe}\mathrm{Te}_{1-x}\mathrm{Se}_x$ \cite{FHsu},  a nesting based approach fails to describe the magnetic ground state \cite{FMa, CFang1, OLipscombe}. These materials have been shown to exhibit both commensurate and incommensurate antiferromagnetic state with a large magnetic moment $\sim 2.0 \mu_B$. However it should be noted that in spite of the stronger correlations in 11-chalcogenides in comparison to pnictides, the parent compound $\mathrm{Fe}\mathrm{Te}$ is still metallic.

The discovery of superconductivity in the new 122 iron-chalcogenides $(\mathrm{K}, \ \mathrm{Tl}, \ \mathrm{Cs}, \ \mathrm{Rb})_y\mathrm{Fe}_{x}\mathrm{Se}_2$ \cite{JGuo, MFang, Krzton-Maziopa, Mizuguchi},
in the vicinity of an antiferromagnetic state with a $T_c \sim 30K$, promises to shed new light on both the role of electronic correlations and nature of magnetism and superconductivity.
In these 122 iron-chalcogenides, there are parent compounds which are antiferromagnetically ordered and insulating \cite{MFang,DMWang}. Furthermore, both the angle resolved photoemission experiments \cite{YZhang, TQian, DMou}
and  band structure calculations  \cite{Shein,XWYan, CChao, LZhang}
reveal unique fermiology. Unlike the pnictides, there are no hole Fermi pockets; the Fermi surface is entirely made up of electron Fermi pockets, with the dominant pieces  located near the  M points of the extended Brillouin zone. The bare magnetic susceptibility $\chi_0(\mathbf{q},\omega)$ for these new materials, as determined from the band structure, is weakly peaked around $\mathbf{q}=(\pi, \pi)$, and can not explain the origin of the observed complex magnetic order with a large magnetic moment ranging from $2 \mu_B$ to $\sim 3.4 \mu_B$ which is intertwined with iron vacancy order \cite{WBao1, WBao2, Yeetal}. The observation of an insulating state and the emergence of antiferromagnetic order with large moment in the absence of quasi-nested electron and hole pockets certainly point towards stronger electronic correlation effects. The possible band narrowing effects due to vacancy ordering, and a resulting Mott insulating phase for the parent materials has been
discussed \cite{RYu1,YZhou}.

In this paper, we study the magnetic order and dynamics of an extended $J_1-J_2$ model on modulated square lattices.
For a lattice (called $L_1$ below) appropriate \cite{MFang}
for the vacancy ordering of a $\mathrm{K}_{y}\mathrm{Fe}_{1.6}\mathrm{Se}_2$, our analysis builds on the important insights introduced by
Cao and Dai \cite{Cao_Dai_block_spin} that vacancy ordering may modulate the exchange interactions.
We show that
the experimentally observed block-spin state
\cite{WBao1, WBao2, Yeetal}
occupies a significant portion of the phase diagram, and we discuss the spin dynamics of this phase using a linear spin-wave analysis. The large unit cell with multiple iron ions leads to both acoustic spin wave and gapped optical spin wave modes. Based on our spin wave calculations in different parameter regimes corresponding to the block spin antiferromagnetic phase, we show how the measurements of spin gaps and degeneracy along the high symmetry directions of the magnetic Brillouin zone provide valuable information regarding the signs of the underlying exchange couplings. We have also analyzed the magnetic phase diagram of a $J_1-J_2$ model on two different modulated square lattices (called $L_2$ and $L_3$
below) potentially
relevant \cite{MFang,ZWang}
 to $\mathrm{K}_y\mathrm{Fe}_{1.5}\mathrm{Se}_2$, which respectively exhibit
 $\frac{1}{4}$-depleted $2\times2$ and $4\times2$ vacancy ordering.
We note that we have not considered the further-neighbor coupling $J_3$,
since the $J_1-J_2$ couplings appear to be adequate for
 $\mathrm{K}_{y}\mathrm{Fe}_{x}\mathrm{Se}_2$.

Our paper is organized as follows. In Sec. II we introduce an extended $J_1-J_2$ model, as appropriate for the modulated square lattice $L_1$ with $\sqrt{5}\times \sqrt{5}$ vacancy order (Fig.~\ref{fig:1a}). In Subsec. II A and Subsec. II B we respectively discuss possible magnetic phases and phase diagram of the extended $J_1-J_2$ model for the $L_1$ lattice, based on a Monte Carlo calculation using classical spins. In Sec. III we discuss the spin wave results for the block spin antiferromagnetic phase. In Sec. IV we introduce the appropriate $J_1-J_2$ models for the modulated lattices $L_2$ (Fig.~\ref{fig:5a}) and $L_3$ (Fig.~\ref{fig:5b}). We discuss possible magnetic phases and phase diagrams for the $L_2$ and $L_3$ lattices, respectively in Subsec. IV A and Subsec. IV B. A summary is given in Sec. V.

\section{Extended $J_1-J_2$ model for $L_1$ lattice with $\sqrt{5}\times \sqrt{5}$ vacancy order} 
\begin{figure}[htbp]
\centering
\subfigure[]{
\includegraphics[scale=0.35,
bbllx=120pt,bblly=-26pt,bburx=591pt,bbury=618pt
]{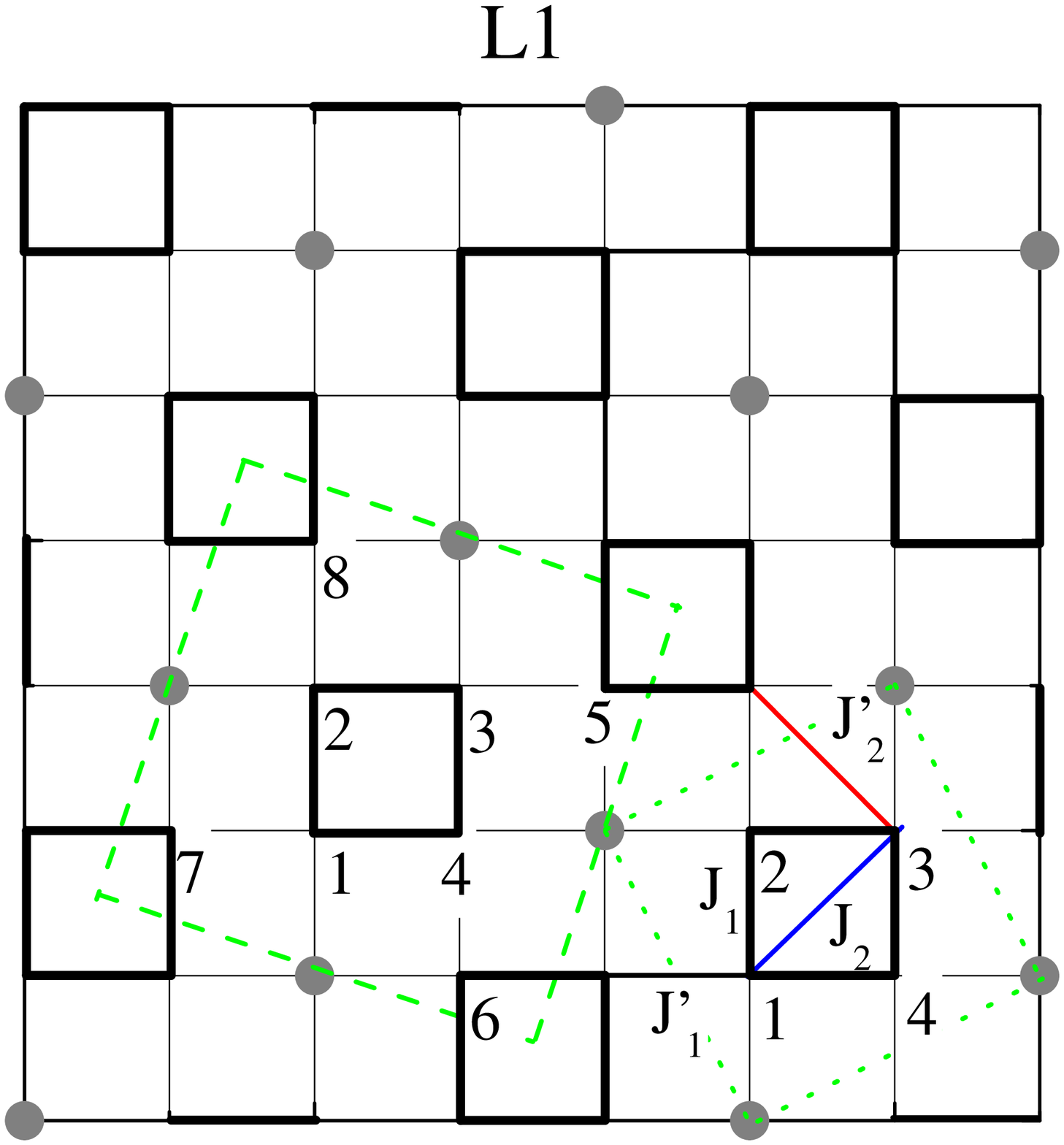}
\label{fig:1a}
}
\subfigure[]{
\includegraphics[scale=0.3,
bbllx=30pt,bblly=20pt,bburx=601pt,bbury=584pt
]{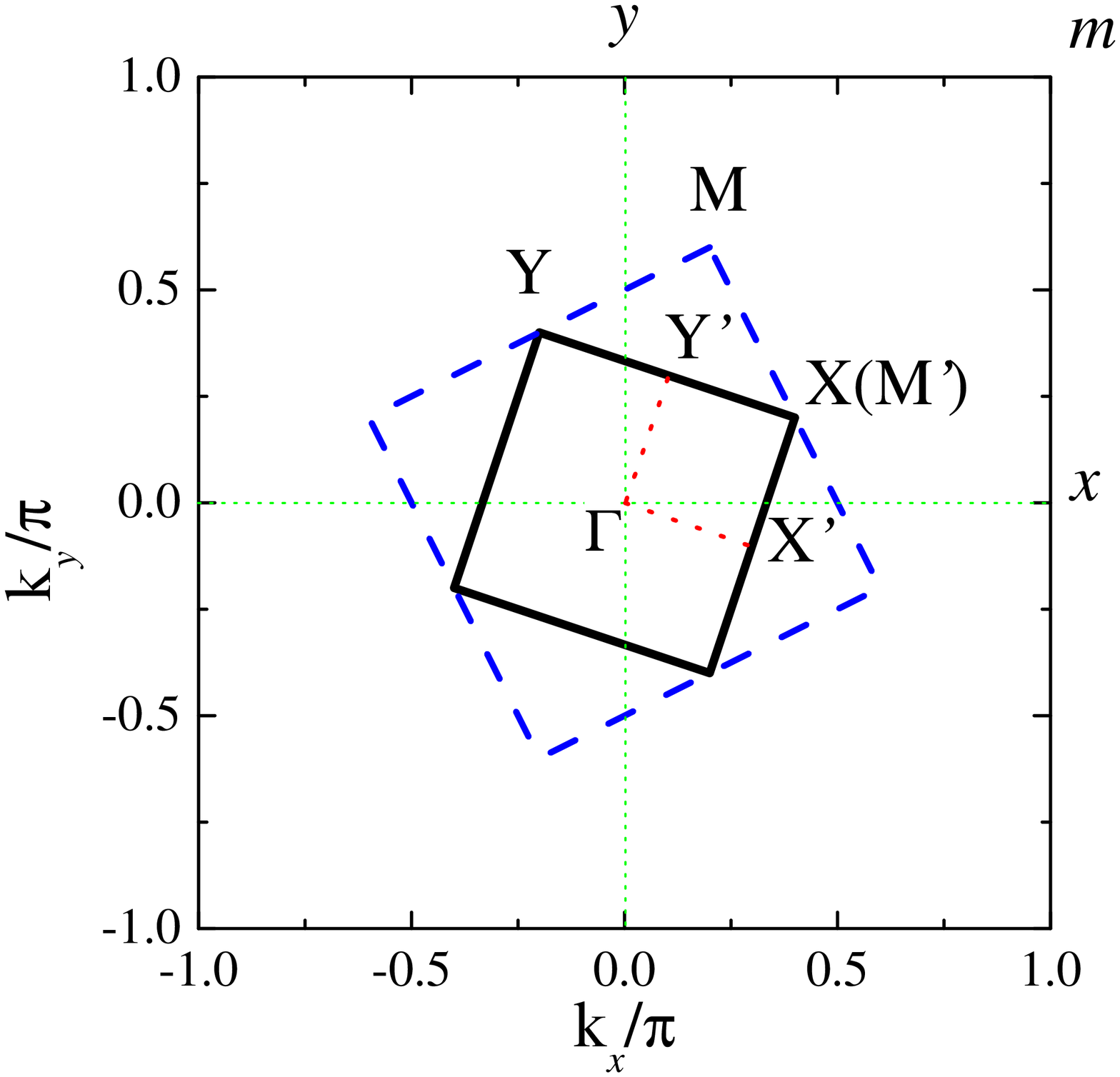}
\label{fig:1b}
}
\label{fig:1}
\caption[]{ In panel (a) we show the $L_1$ lattice with $\sqrt{5}\times \sqrt{5}$ vacancy order, with the shaded circles describing the vacancies. The structural unit cell due to the vacancy ordering is marked by dotted green lines, with four $\mathrm{Fe}$ sites residing on a plaquette $1234$ which is marked by thick, dark lines. The intra unit cell exchange couplings $J_1$, $J_2$ and inter unit cell exchange coupling $J_{1}^{\prime}$, $J_{2}^{\prime}$ are also illustrated. The larger unit cell consisting of eight sites $1,2,...,8$, and marked by dashed green lines corresponds to the magnetic unit cell of the AF1 phase. In panel (b) we show three different Brillouin zones (BZ). The extended BZ (EBZ) $-\pi \leq k_x \leq \pi$, $-\pi \leq k_x \leq \pi$, marked by solid black lines corresponds to 122-unit cell that encloses one $\mathrm{Fe}$ site. The structural BZ (SBZ) of the modulated lattice corresponding to the unit cell with four sites is marked by the dashed blue lines. This BZ encloses $\frac{1}{5}$-th of the area enclosed by the EBZ. The smallest BZ also marked by solid black lines corresponds to the magnetic BZ of the AF1 phase (MBZ1). The high-symmetry points in the SBZ and MBZ1 have the following coordinates: in SBZ, $\mathrm{X}=(\frac{2\pi}{5},\frac{\pi}{5})$, $\mathrm{Y}=(-\frac{\pi}{5},\frac{2\pi}{5})$, $\mathrm{M}=(\frac{\pi}{5},\frac{3\pi}{5})$; and in MBZ1, $\mathrm{X'}=(\frac{3\pi}{10},-\frac{\pi}{10})$, $\mathrm{Y'}=(\frac{\pi}{10},\frac{3\pi}{10})$, $\mathrm{M'}=(\frac{2\pi}{5},\frac{\pi}{5})$.
}
\end{figure}

The modulated lattice $L_1$ with $\sqrt{5} \times \sqrt{5}$ vacancy order is shown in Fig.~\ref{fig:1a}. Shown here is the structural unit cell comprising four $\mathrm{Fe}$ sites and a larger unit cell consisting of eight $\mathrm{Fe}$ sites. The eight site unit cell is the magnetic unit cell of the experimentally relevant block-spin antiferromagnetic state and this also turns out to be convenient to describe other anitiferromagnetic phases to be specified below. In Fig.~\ref{fig:1b} we show the structural Brillouin zone (SBZ) corresponding to the four-site unit cell and the magnetic Brillouin zone of the block-spin antiferromagnetic state (MBZ1) corresponding to the eight site unit cell.
A recent DFT calculation~\cite{Yanetal2011} shows that the Fe vacancy order induces a tetramer lattice distortion which may strongly influence the superexchange couplings. This distortion makes the intra-block and inter-block exchange couplings different, and leads to the following modulated $J_1-J_2$ Hamiltonian~\cite{Cao_Dai_block_spin}
\begin{eqnarray}
H_1 &=& \frac{1}{2}\sum_{i}\bigg \{ J_1 \sum_{\langle \alpha \beta \rangle} \mathbf{S}_{i\alpha}\cdot \mathbf{S}_{i\beta}+ J_2 \sum_{\langle \langle \alpha \beta \rangle \rangle} \mathbf{S}_{i\alpha}\cdot \mathbf{S}_{i\beta}\bigg \} \nonumber \\
&+& \frac{1}{2}\sum_{i\neq j}\bigg \{ J_{1}^{\prime} \sum_{\langle \alpha \beta \rangle} \mathbf{S}_{i\alpha}\cdot \mathbf{S}_{j\beta}+ J_{2}^{\prime} \sum_{\langle \langle \alpha \beta \rangle \rangle} \mathbf{S}_{i\alpha}\cdot \mathbf{S}_{j\beta}\bigg \},
\label{eq:1}
\end{eqnarray}
where the Latin indices $i, j$ correspond to the unit cells, each enclosing a single vacancy and comprising four $\mathrm{Fe}$ sites. The Greek indices $\alpha, \beta =1,2,3,4$ correspond to the four $\mathrm{Fe}$ sites of the unit cell. We have respectively denoted the nearest and next nearest neighbors by $\langle \alpha \beta \rangle$ and $\langle \langle \alpha \beta \rangle \rangle$. Within the same unit cell,
the exchange couplings between the nearest and next nearest neighbor iron sites
are respectively denoted by $J_1$ and $J_2$. Between the neighboring unit cells,
the couplings between the nearest and next nearest neighbor iron sites are
respectively $J_1^{\prime}$ and $J_2^{\prime}$.
In order to incorporate the modulations of the exchange couplings by the vacancy order,
we will allow  the exchange couplings to take both antiferromagnetic ($J_a, \ J_{a}^{\prime} >0$) and ferromagnetic ($J_a, \ J_{a}^{\prime}<0$) signs.

\subsection{Magnetic phases for $L_1$ lattice}
\begin{figure}[htbp]
\centering
\subfigure[]{
\includegraphics[scale=0.18,
bbllx=60pt,bblly=40pt,bburx=621pt,bbury=604pt
]{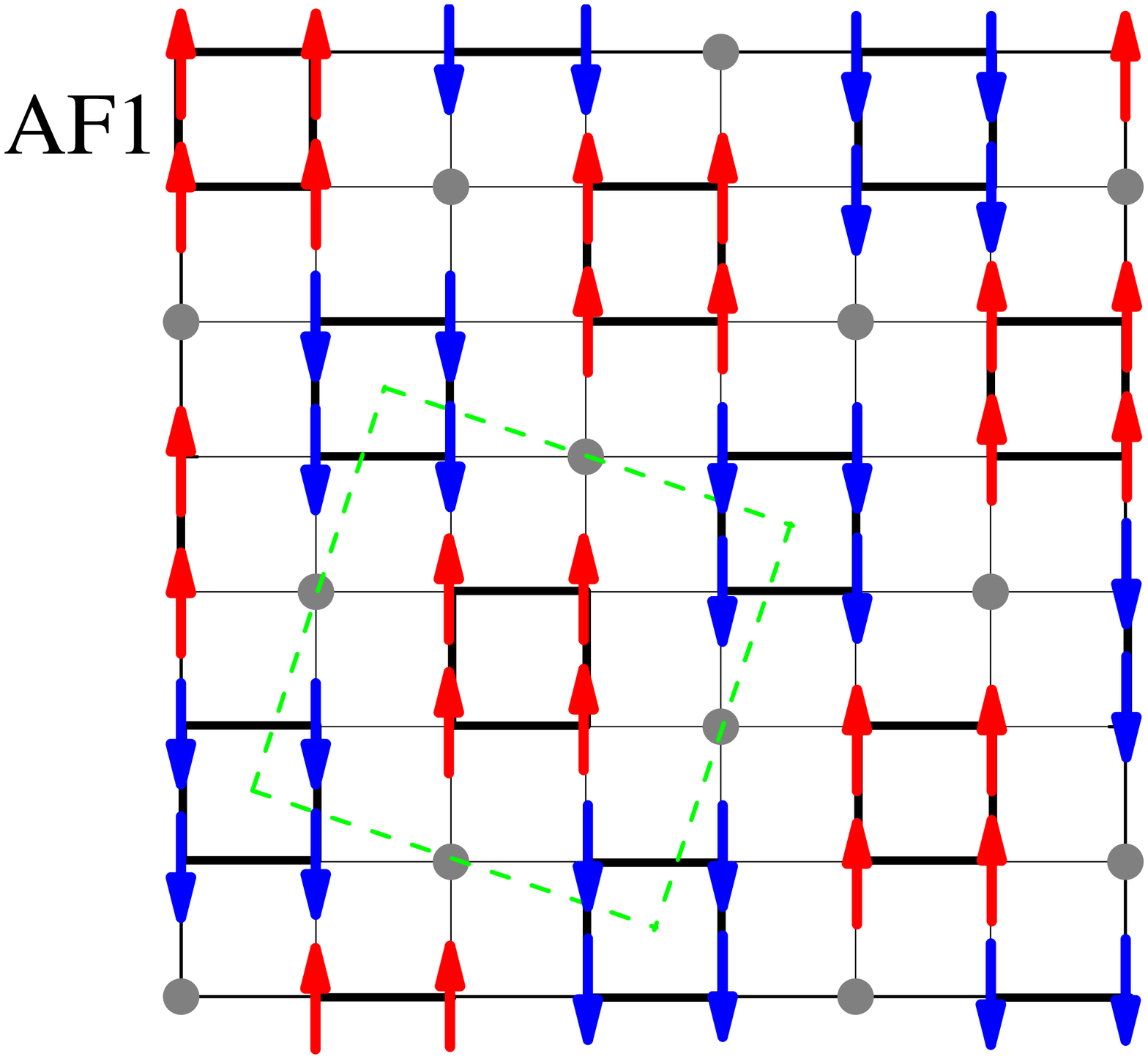}
\label{fig:2a}
}
\subfigure[]{
\includegraphics[scale=0.18,
bbllx=10pt,bblly=40pt,bburx=571pt,bbury=604pt
]{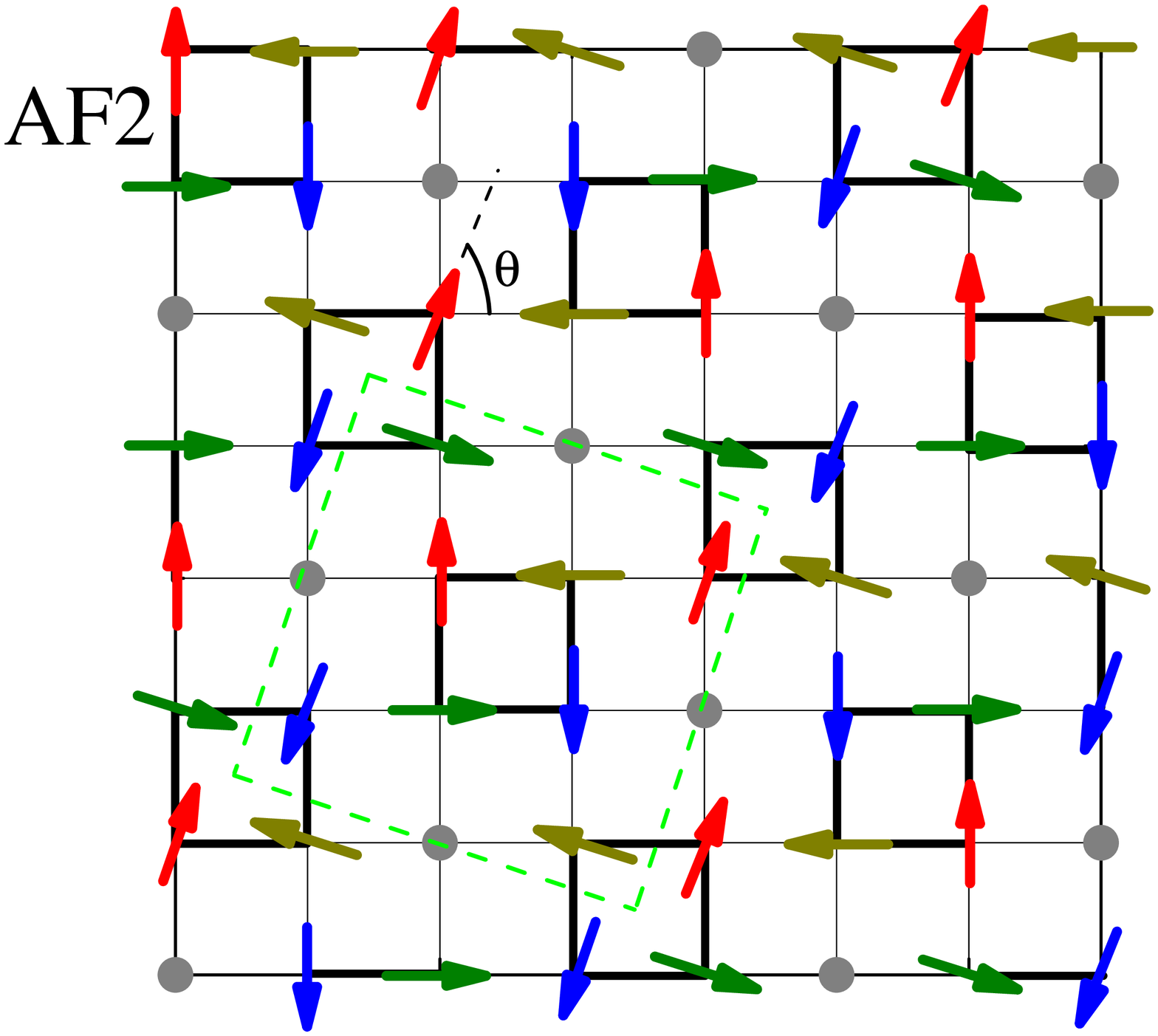}
\label{fig:2b}
}
\subfigure[]{
\includegraphics[scale=0.18,
bbllx=60pt,bblly=40pt,bburx=621pt,bbury=604pt
]{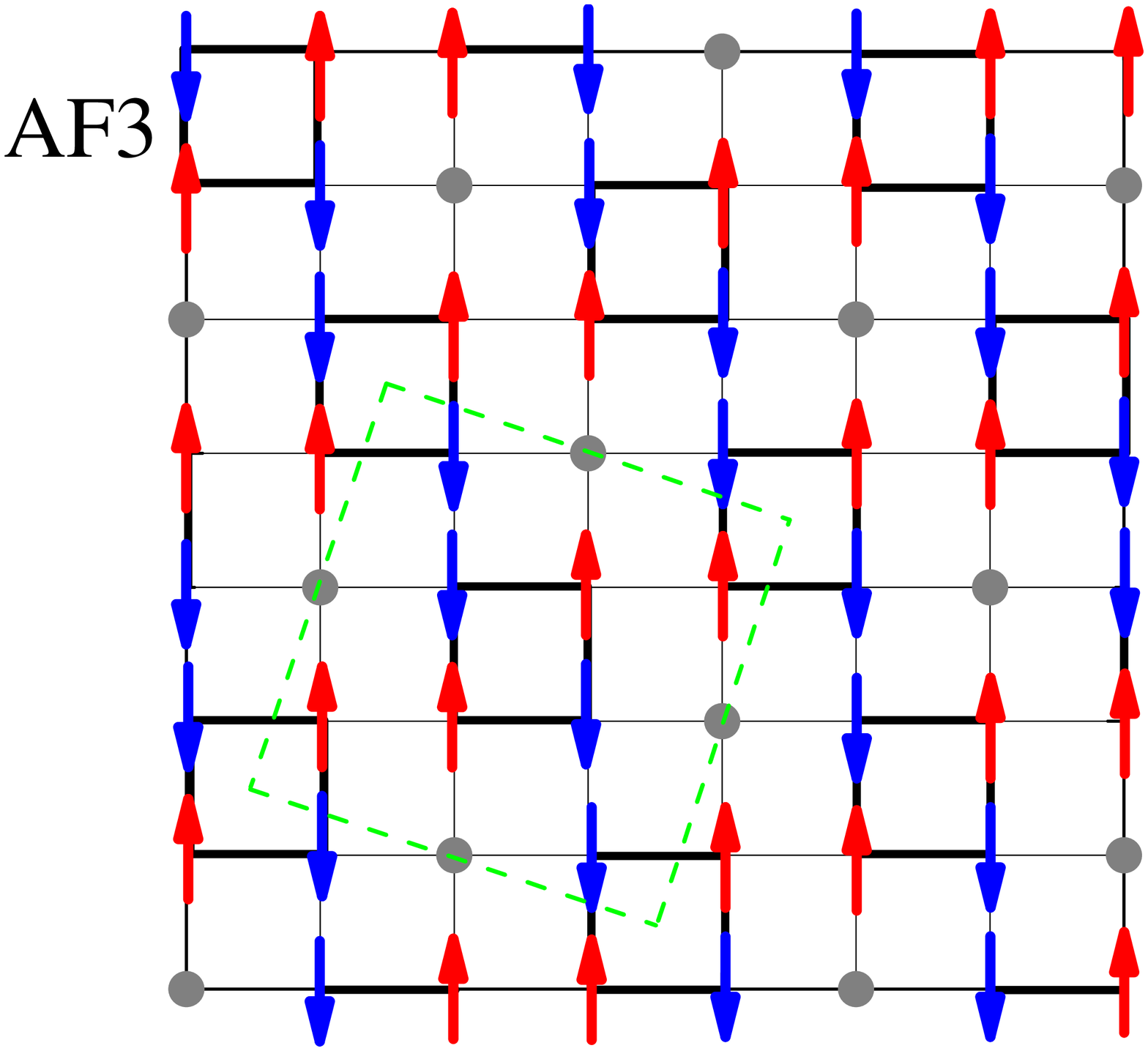}
\label{fig:2c}
}
\subfigure[]{
\includegraphics[scale=0.18,
bbllx=10pt,bblly=40pt,bburx=571pt,bbury=604pt
]{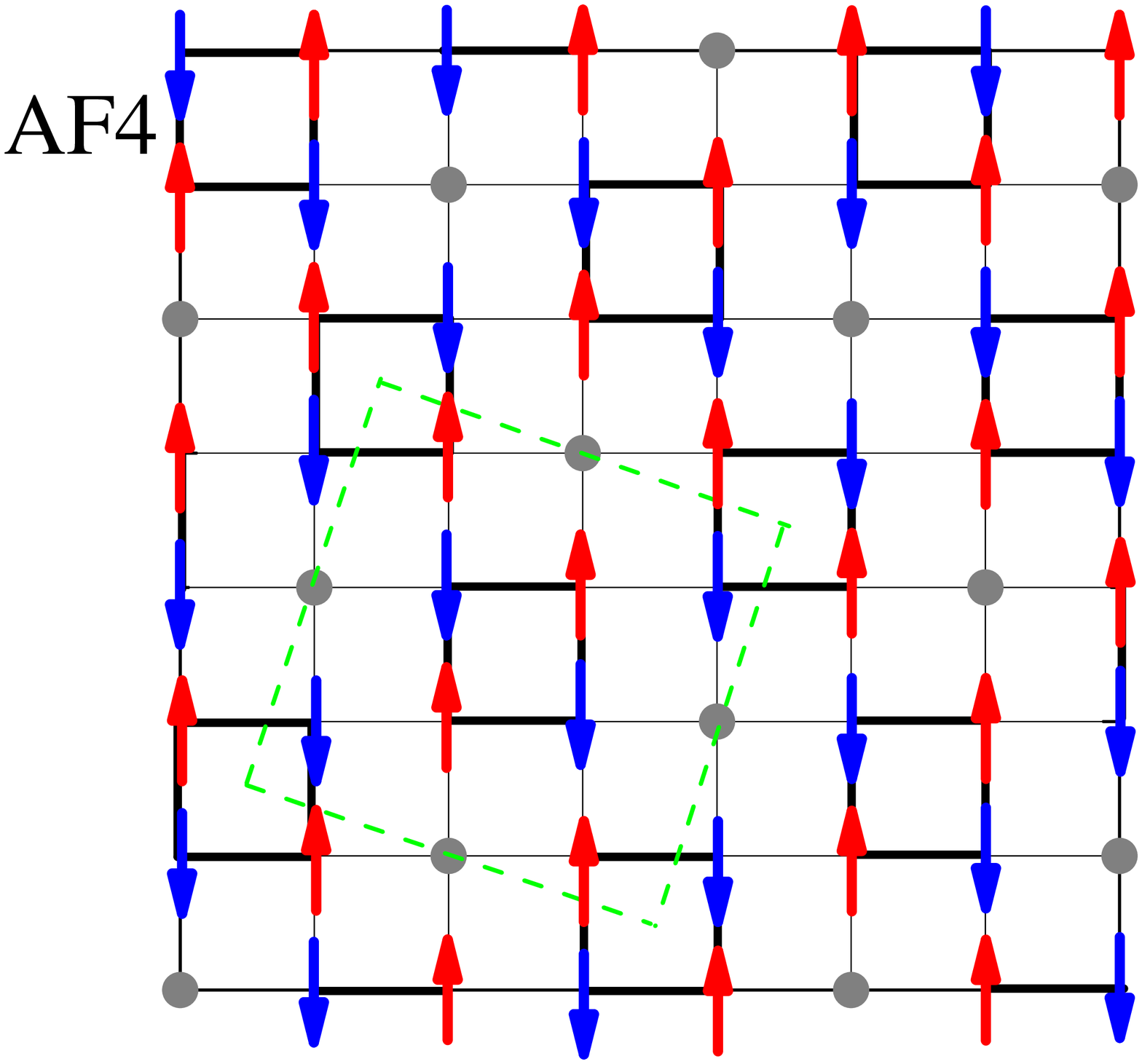}
\label{fig:2d}
}
\subfigure[]{
\includegraphics[scale=0.18,
bbllx=60pt,bblly=40pt,bburx=621pt,bbury=604pt
]{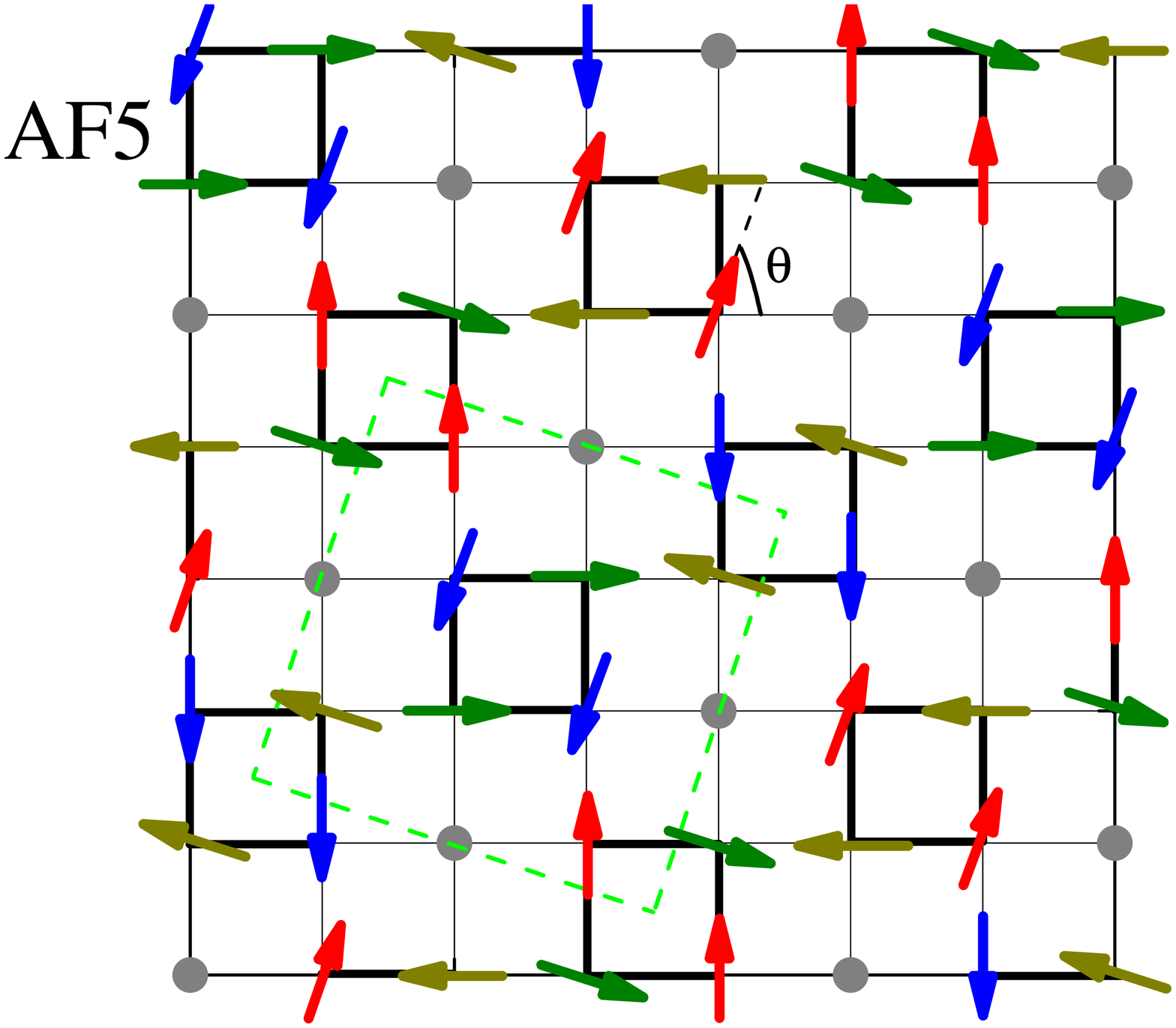}
\label{fig:2e}
}
\subfigure[]{
\includegraphics[scale=0.18,
bbllx=10pt,bblly=40pt,bburx=571pt,bbury=604pt
]{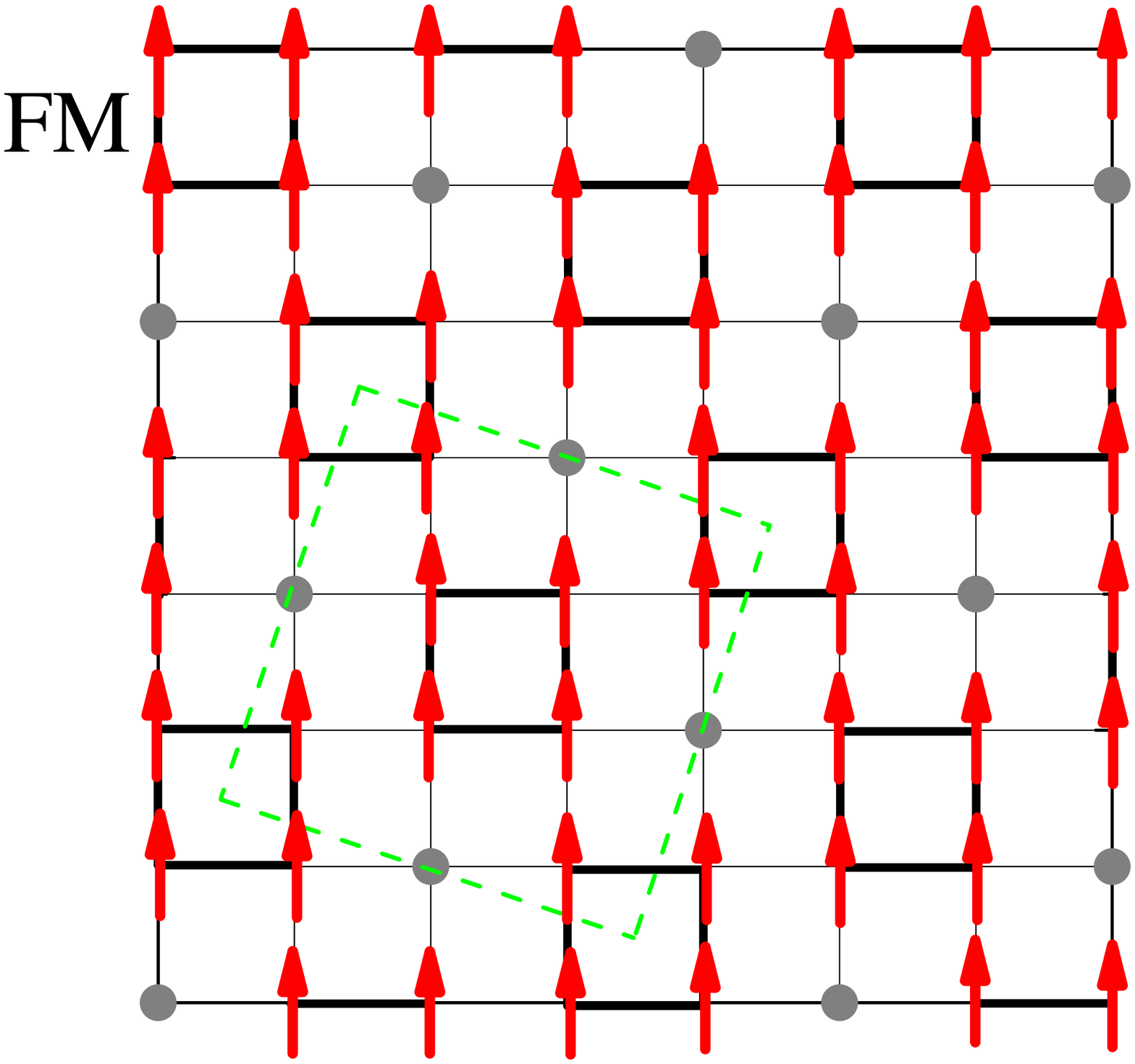}
\label{fig:2f}
}
\label{fig:2}
\caption[]{In panels (a) through (f) we show the real space arrangements of the spins for six commensurate magnetic states. AF1, AF3, AF4 are collinear antiferromagnetic states with ordering vector $\mathbf{Q}=(0,0)$ in the MBZ1, and these phases are distinguished by the different canting angle assignments. AF2 and AF5 are non-collinear antiferromagnetic states, with ordering vector $\mathbf{Q}=(\pi, \pi)$ in the MBZ1. The conventional ferromagnetic state is denoted as FM and has $\mathbf{Q}=(0,0)$. AF1 phase corresponds to the experimentally relevant block spin antiferromagnetic phase.
}
\end{figure}

To determine the phase diagram of the extended $\mathrm{J}_1$-$\mathrm{J}_2$ model defined by Eq.~\ref{eq:1}, we apply classical Monte Carlo technique. We treat the spins as classical $O(3)$ vectors, and adapt the standard Metropolis algorithm. The Monte Carlo simulations have been performed on finite lattices with linear dimension up to $80$, and at temperatures as low as $T\sim10^{-3}J_2^\prime$. By checking both the static spin structure factor and the Monte Carlo snapshots of spins in real space, we identify the ordering wave vectors and corresponding real-space spin pattern of various phases stabilized at low temperatures. We have found seven possible magnetic phases. One of these is an incommensurate magnetic phase (no higher order commensuration), with very broad structure factor in the entire EBZ, SBZ and MBZ1. This phase occupies a small sliver of the parameter space in the phase diagram and will not be discussed in detail. In the magnetic phase diagram we denote this incommensurate phase as IC. The other six magnetically ordered phases have commensurate ordering vectors in the MBZ1. These phases are denoted as AF1, AF2, AF3, AF4, AF5 and FM in the following discussions, and are shown in Fig.~\ref{fig:2a} through Fig.~\ref{fig:2f}. The AF1 phase corresponds to the experimentally relevant block-spin antiferromagnetic state.

All the six magnetic states shown in Fig. 2 can be described by the
following single-mode spiral:
$\mathbf{S}_{i\alpha} = \hat{x}\cos(\mathbf{Q}\cdot\mathbf{R}_i+\theta_\alpha)
+ \hat{y}\sin(\mathbf{Q}\cdot\mathbf{R}_i+\theta_\alpha)$, where
$\mathbf{R}_i$ and $\mathbf{Q}$ are respectively the position vector
in the 8-site unit cell, and the ordering wave vector in the MBZ1.
The canting angle at site $\alpha$ is denoted by $\theta_\alpha$.
The fact that these six states are the lowest energy states, has also
been confirmed by minimizing the ground-state energy with respect to
$\mathbf{Q}$ and $\theta_\alpha$. The commensurate ordering wave vectors,
the canting angles and the classical ground state energy per site are
listed in TABLE. I.

\begin{center}
\begin{table}[htdp]
\begin{tabular}{|c|c|p{2.5cm}|c|}
\hline
Phase & $\mathbf{Q}$-vector & $\quad\quad\quad\theta_{\alpha}$ & Energy per site \tabularnewline
\hline
\cline{1-4}
AF1 & $\mathbf{Q}=(0,0)$ & $\theta_{1,2,3,4}=0$ & $J_1-J'_2+\frac{J_2-J'_1}{2}$\\
 & & $\theta_{5,6,7,8}=\pi$&
\tabularnewline
\hline
AF2 & $\mathbf{Q}=(\pi,\pi)$ & $\theta_{i\leqslant4}=\frac{(i-1)\pi}{2}$ & $-\frac{J_2}{2}-\sqrt{\frac{J'^2_1}{4}+J'^2_2}$\\
& & $\theta_{i\geqslant5} = \frac{(i-5)\pi}{2} + \tan^{-1}\left(\frac{2J'_2}{J'_1}\right)$&
\tabularnewline
\hline
AF3 & $\mathbf{Q}=(0,0)$ & $\theta_{1,3,5,7}=0$& $\frac{J_2+J'_1}{2}-J_1-J'_2$\\
& & $\theta_{2,4,6,8}=\pi$&
\tabularnewline
\hline
AF4 & $\mathbf{Q}=(0,0)$ & $\theta_{1,3,6,8}=0$& $\frac{J_2-J'_1}{2}-J_1+J'_2$\\
& & $\theta_{2,4,5,7}=\pi$&
\tabularnewline
\hline
AF5 & $\mathbf{Q}=(\pi,\pi)$ & $\theta_{1,3}=0$, $\theta_{6,8}=\frac{\pi}{2}$ & $\frac{J_2}{2}-\sqrt{\frac{J'^2_1}{4}+J^2_1}$\\
& & $\theta_{2,4}=\pi+\theta_0$&\\
& & $\theta_{5,7}=\pi/2+\theta_0$&\\
& & $\theta_0=\tan^{-1}\left(\frac{J'_1}{2J_1}\right)$&
\tabularnewline
\hline
FM & $\mathbf{Q}=(0,0)$ & $\theta_{i}=0$ for & $J_1+J'_2+\frac{J_2+J'_1}{2}$\\
& & $i=1,...,8$ &\\
\hline
\end{tabular}
\caption{The ordering wave vector $\mathbf{Q}$ in MBZ1, the canting angles and the ground-state energies per Fe site of the six commensurate magnetic states shown in Fig.~\ref{fig:2a} through Fig.~\ref{fig:2f}.}\label{tab:1}
\end{table}
\end{center}

The four phases AF1, AF3, AF4 and FM are described by the same ordering wavevector $\mathbf{Q}=(0,0)$ in the MBZ1, whereas $\mathbf{Q}=(\pi,\pi)$ for AF2 and AF5. The phases with same ordering vectors are distinguished by the assignment of the canting angle values. Depending on the canting angle values, we obtain collinear and non-collinear magnetic states. The AF1 phase is the collinear, block-spin antiferromagnetic phase found in the neutron diffraction experiments on $\mathrm{K}_{0.8}\mathrm{Fe}_{1.6}\mathrm{Se}_2$. On a four site unit cell all the spins are ferromagnetically aligned and form a large block spin, and these block spins on the adjacent four site unit cells are antiferromagnetically aligned. This state can be equivalently described as a $(\pi, \pi)$ AF ordered phase in the SBZ. In the following section we address the role of quantum fluctuations in this phase using linear spin wave analysis.

The AF2 phase is a non-collinear antiferromagnetic state. On each four site unit cell, nearest neighbor spins are orthogonal to each other, whereas the next nearest neighbor spins are antiparallel to each other. The canting angle for all the spins on the adjacent unit cells are shifted by an angle  $\mathrm{arctan}(2J_{2}^{\prime}/J_{1}^{\prime})$. In terms of the SBZ, the ordering wavevector is at both $(\pi,0)$ and $(0,\pi)$. Note that this state is different from the collinear $(\pi,0)$/$(0,\pi)$ state in the $J_1-J_2$ model when $J_2\gtrsim J_1/2$. The AF2 state does not break the $C_4$ rotational symmetry, and it cannot be described by a single ordering wavevector in the SBZ. The AF3 phase is a collinear magnetic state. On a four site unit cell the spins are antiferromagnetically aligned and each unit cell has the same spin arrangement. Therefore in the SBZ this state has ordering vector $(0,0)$. In the AF4 phase every neighboring spin is antiferromagnetically aligned and this phase corresponds to conventional Neel state on the 122 unit cell. Also notice that on each four site unit cell spins are antiferromagnetically aligned, but on the adjacent unit cells all the spins are flipped. Therefore in terms of SBZ AF4 has $(\pi, \pi)$ ordering vector. The AF5 phase exhibits non-collinear magnetic order. On each four site unit cell the canting between nearest neighbor spins depends on the ratio $J_{1}^{\prime}/2J_{1}$, whereas the next nearest neighbors are ferromagnetically aligned. Going to an adjacent unit cell all the canting angles are shifted by $\pi/2$. Finally the FM phase corresponds to the conventional ferromagnetic state. In the following subsection we describe the associated magnetic phase diagram.

\subsection{Phase diagram for $L_1$ lattice}
\begin{figure}[htbp]
\centering
\subfigure[]{
\includegraphics[scale=0.2,
bbllx=80pt,bblly=10pt,bburx=641pt,bbury=564pt
]{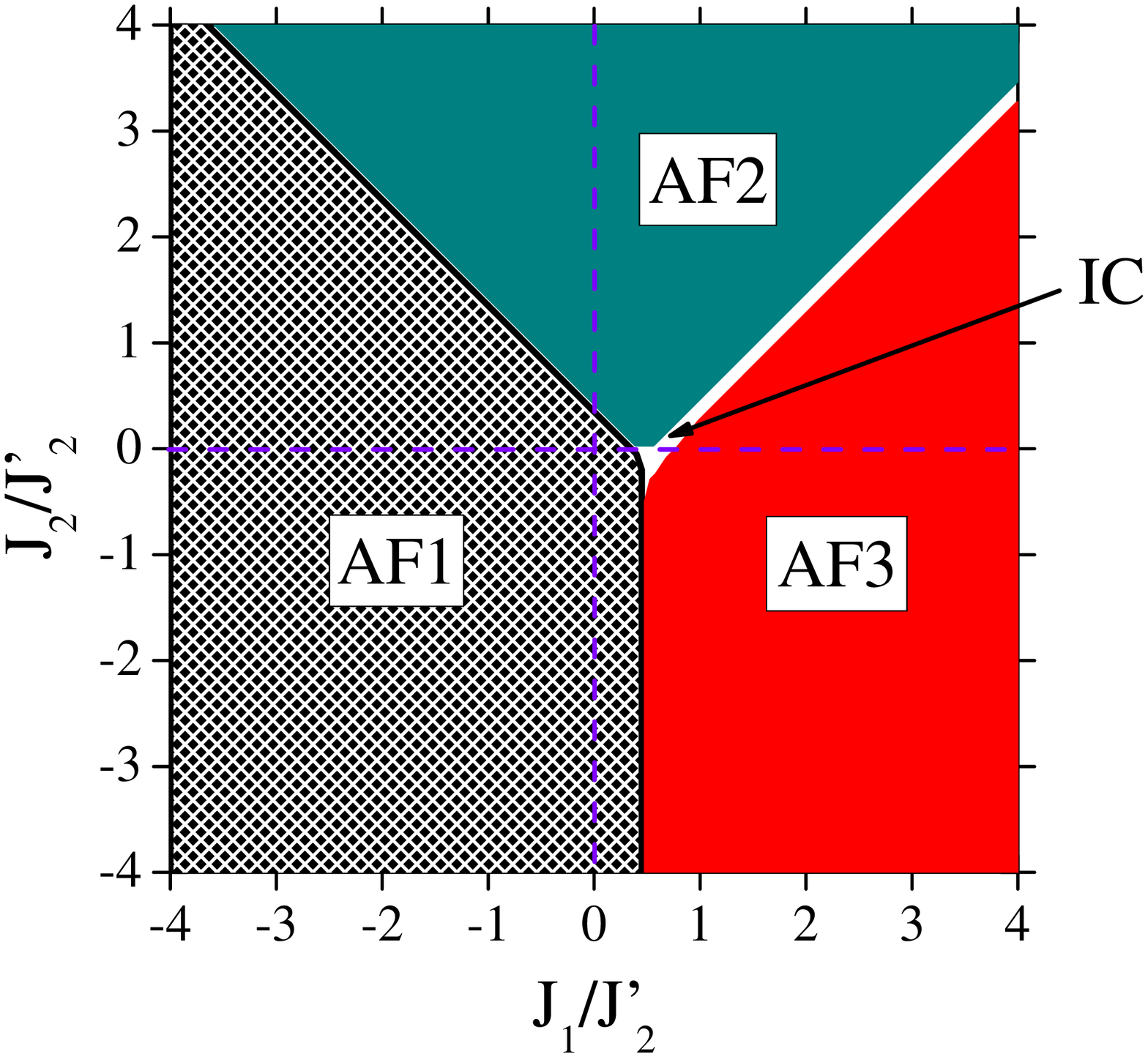}
\label{fig:3a}
}
\subfigure[]{
\includegraphics[scale=0.2,
bbllx=40pt,bblly=10pt,bburx=601pt,bbury=564pt
]{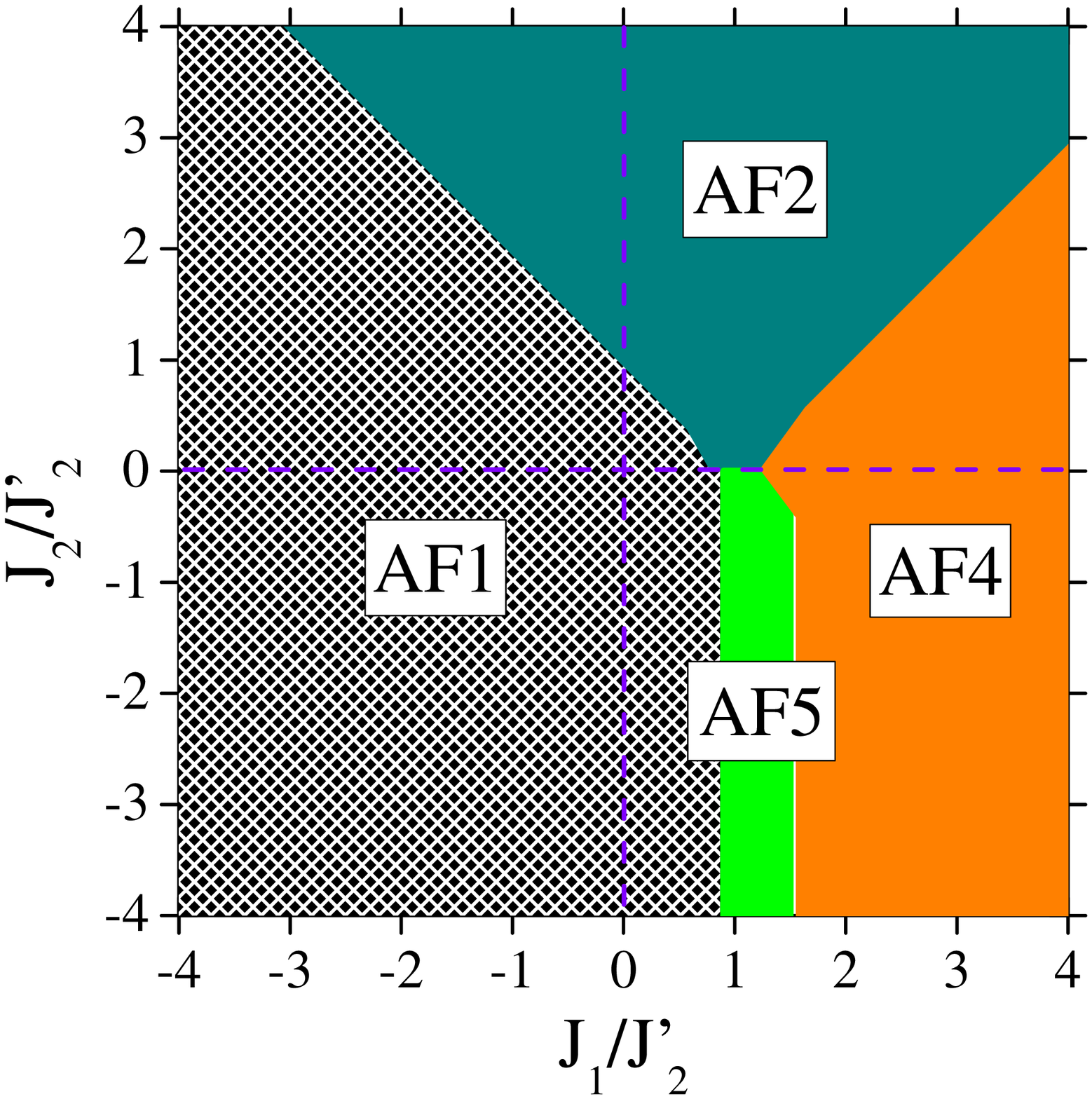}
\label{fig:3b}
}
\subfigure[]{
\includegraphics[scale=0.2,
bbllx=70pt,bblly=10pt,bburx=631pt,bbury=564pt
]{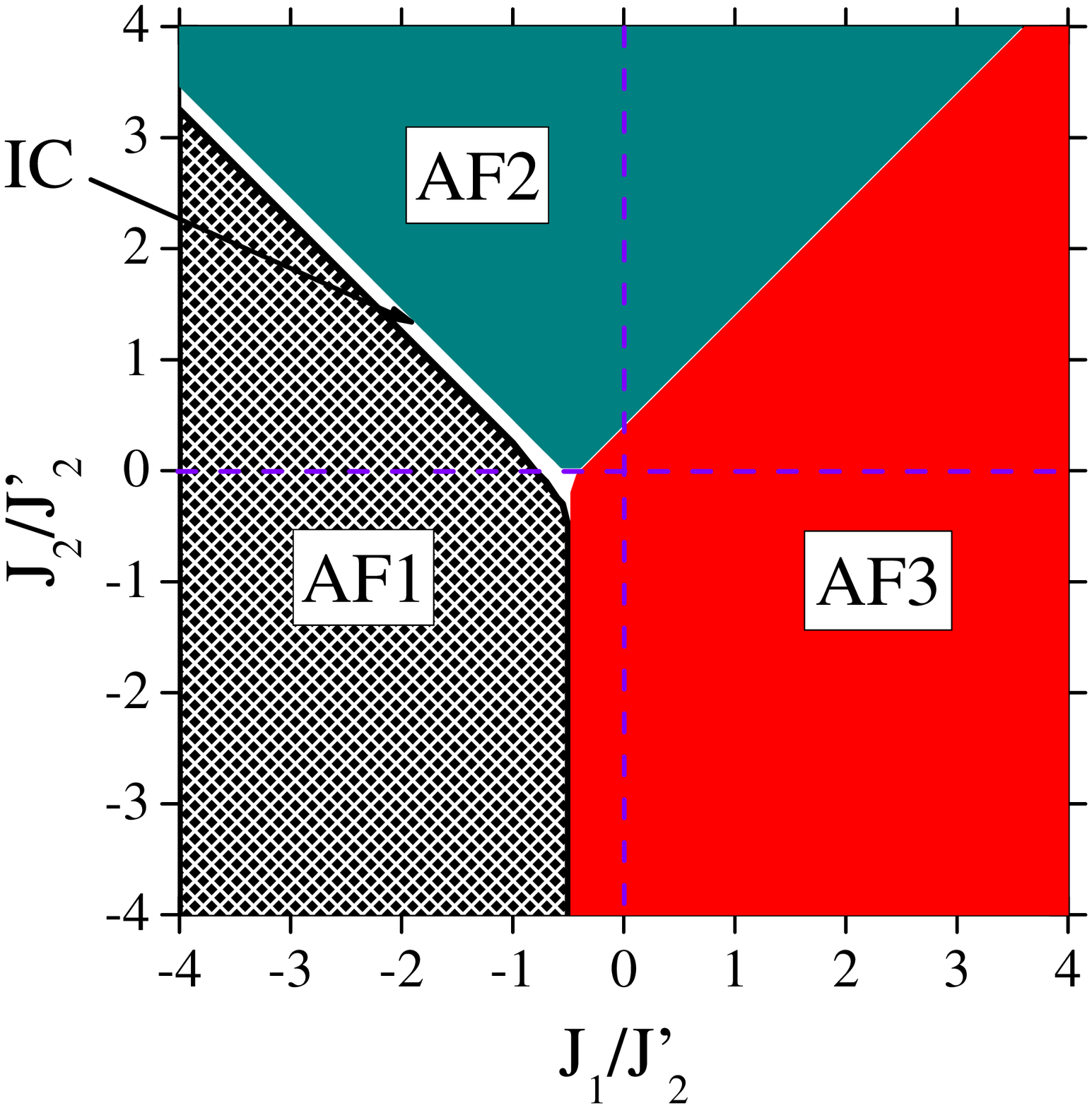}
\label{fig:3c}
}
\subfigure[]{
\includegraphics[scale=0.2,
bbllx=40pt,bblly=10pt,bburx=601pt,bbury=564pt
]{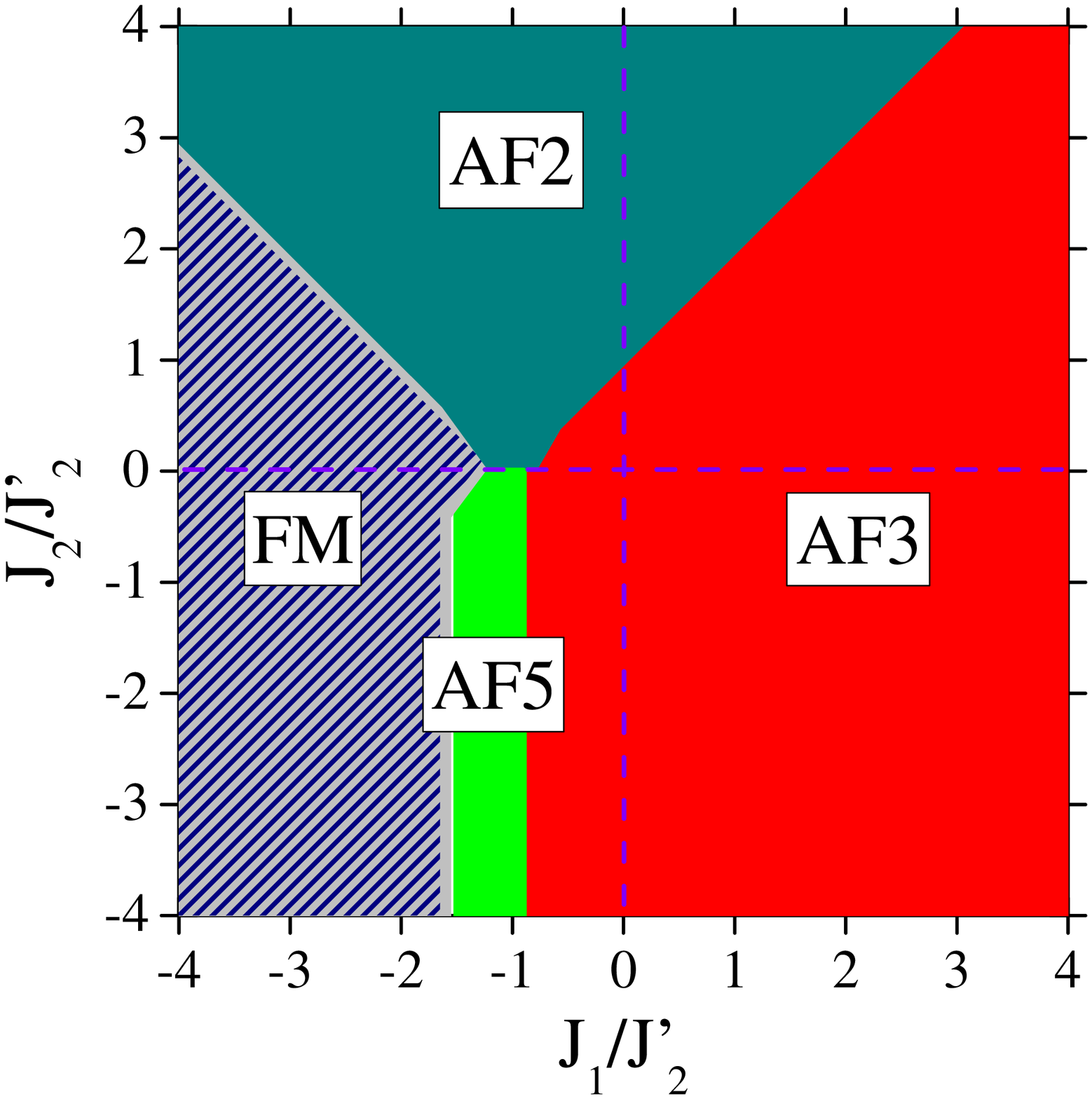}
\label{fig:3d}
}
\label{fig:3}
\caption[]{In panels (a) through (d) we respectively show the magnetic phase diagrams in the $J_1/J_{2}^{\prime}-J_2/J_{2}^{\prime}$ plane for $J_{1}^{\prime}/J_{2}^{\prime}=1, \ 4, \ -1, \ -4$. The phase diagrams in panels (a), (b), (c) and (d) are respectively applicable
for $J_{1}^{\prime}/J_{2}^{\prime}<2$, $J_{1}^{\prime}/J_{2}^{\prime}>2$, $J_{1}^{\prime}/J_{2}^{\prime}>-2$, $J_{1}^{\prime}/J_{2}^{\prime}<-2$, with only quantitative shifts of the phase boundaries.}
\end{figure}

The seven phases mentioned above give rise to a complex phase diagram, and most of the phase diagram excluding the incommensurate IC phase, can be obtained by comparing the tabulated energies for the six magnetic phases. We measure energies in the units of $J_2^{\prime}$ and in Fig.~\ref{fig:3a} through Fig.~\ref{fig:3d} we plot the phase diagrams in the $J_1/J_{2}^{\prime}-J_2/J_{2}^{\prime}$ plane, for four different values of $J_{1}^{\prime}/J_{2}^{\prime}$ ratio. The phase diagram in Fig.~\ref{fig:3a} corresponds to $J_{1}^{\prime}/J_{2}^{\prime}=1$ and qualitatively remains applicable for $J_{1}^{\prime}/J_{2}^{\prime}<2$. The phase diagram in Fig.~\ref{fig:3b} corresponds to $J_{1}^{\prime}/J_{2}^{\prime}=4$ and is qualitatively applicable for any $J_{1}^{\prime}/J_{2}^{\prime}>2$. The phase diagram in Fig.~\ref{fig:3c} corresponds to $J_{1}^{\prime}/J_{2}^{\prime}=-1$ and is qualitatively applicable for any $J_{1}^{\prime}/J_{2}^{\prime}>-2$. The phase diagram in Fig.~\ref{fig:3d} corresponds to $J_{1}^{\prime}/J_{2}^{\prime}=-4$ and is qualitatively applicable for any $J_{1}^{\prime}/J_{2}^{\prime}<-2$. Considering the symmetry of the Hamiltonian in Eq.~\ref{eq:1}, we find the ground states for $J^\prime_1<0$ or $J^\prime_2<0$ can be obtained from the states in the phase diagram at $J^\prime_1>0$ and $J^\prime_2>0$ by performing the following two transformations: a) sending $\mathbf{S}_\alpha\rightarrow -\mathbf{S}_\alpha$ for $\alpha=2,4,5,7$ in the 8-site unit cell, and $J_1\rightarrow -J_1$, $J^\prime_1\rightarrow -J^\prime_1$; b) sending $\mathbf{S}_\alpha\rightarrow -\mathbf{S}_\alpha$ for $\alpha=2,4,6,8$, and $J_1\rightarrow -J_1$, $J^\prime_2\rightarrow -J^\prime_2$. The experimentally relevant block spin state AF1 occupies a large portion of the phase diagram. Only for $J_{1}^{\prime}/J_{2}^{\prime}<-2$ is the AF1 phase absent. This provides a simple constraint on the $J_{1}^{\prime}/J_{2}^{\prime}$ ratio. It is important to note that AF1 phase extends into a small region where all the exchange couplings are antiferromagnetic. We further note that the values of the exchange couplings predicted in Ref.~\onlinecite{Cao_Dai_block_spin} based on LDA calculations are close to the AF1-AF2 phase boundary. This suggests, on one hand, it is possible that in certain materials $J_2$ can be tuned to cross the AF1-AF2 phase boundary to stabilize the AF2 ground state, which can still be described by our theory. On the other hand, it also suggests considerable quantum fluctuations in the experimentally observed block-spin (AF1) state. In the next section we investigate the quantum fluctuations in the AF1 state within the framework of linear spin-wave theory. We demonstrate how the spin wave dispersions in the AF1 phase depend on the underlying exchange constants in the three regimes corresponding to Fig.~\ref{fig:3a} through Fig.~\ref{fig:3c}.

\section{Spin wave results for block spin antiferromagnetic state AF1}
To study the effects of quantum fluctuations on the block-spin antiferromagnetic state using linear spin-wave theory, we introduce the following linearized Holstein-Primakoff (HP) transformations. Within the 8-site unit cell, we have $S^z_{i\alpha}=S-a^\dagger_{i\alpha}a_{i\alpha}$, $S^+_{i\alpha}=\sqrt{2S}a_{i\alpha}$, $S^-_{i\alpha}=\sqrt{2S}a^\dagger_{i\alpha}$ for $\alpha=1,2,3,4$, and $S^z_{i\alpha}=-S+a^\dagger_{i\alpha}a_{i\alpha}$, $S^+_{i\alpha}=\sqrt{2S}a^\dagger_{i\alpha}$, $S^-_{i\alpha}=\sqrt{2S}a_{i\alpha}$ for $\alpha=5,6,7,8$. After performing the Fourier transformation, and defining the spinor $\psi^\dagger_\mathbf{k} = \left( a^\dagger_{1,\mathbf{k}},a^\dagger_{2,\mathbf{k}},a^\dagger_{3,\mathbf{k}}, a^\dagger_{4,\mathbf{k}},a_{5,-\mathbf{k}},a_{6,-\mathbf{k}}, a_{7,-\mathbf{k}},a_{8,-\mathbf{k}}\right)$, the Hamiltonian $H=\sum_{\mathbf{k}} \psi^\dagger_\mathbf{k} \mathcal{H_\mathbf{k}} \psi_\mathbf{k}$ can be diagonalized via a Bogoliubov transformation $b^\dagger_{\alpha,\mathbf{k}} = \sum_{\beta=1,2,3,4} U_{\alpha\beta,\mathbf{k}} a^\dagger_{\beta,\mathbf{k}} + V_{\alpha\beta,\mathbf{k}} a_{4+\beta, -\mathbf{k}}$, where $\sum_{\beta} (|V_{\alpha\beta,\mathbf{k}}|^2 - |U_{\alpha\beta,\mathbf{k}}|^2)=1$.

\begin{figure}[htbp]
\centering
\subfigure[]{
\includegraphics[scale=0.6,
bbllx=180pt,bblly=70pt,bburx=340pt,bbury=270pt
]{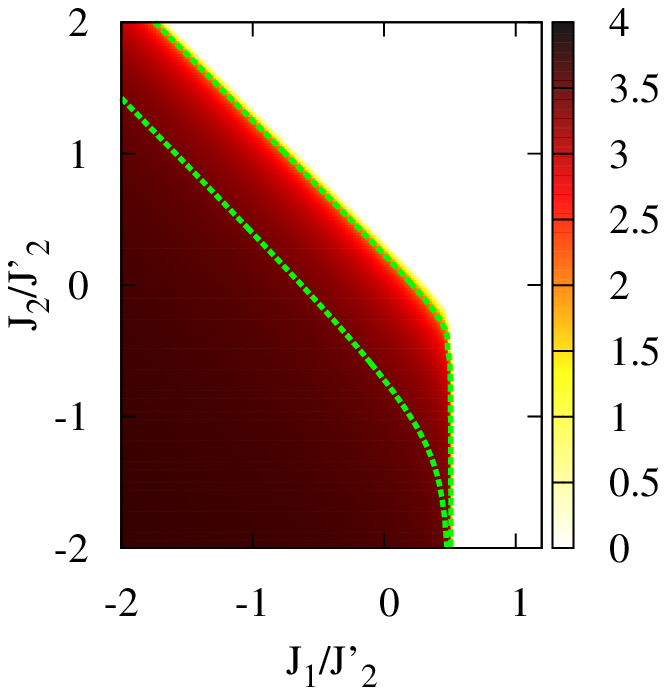}
\label{fig:8a}
}
\subfigure[]{
\includegraphics[scale=0.6,
bbllx=130pt,bblly=70pt,bburx=290pt,bbury=270pt
]{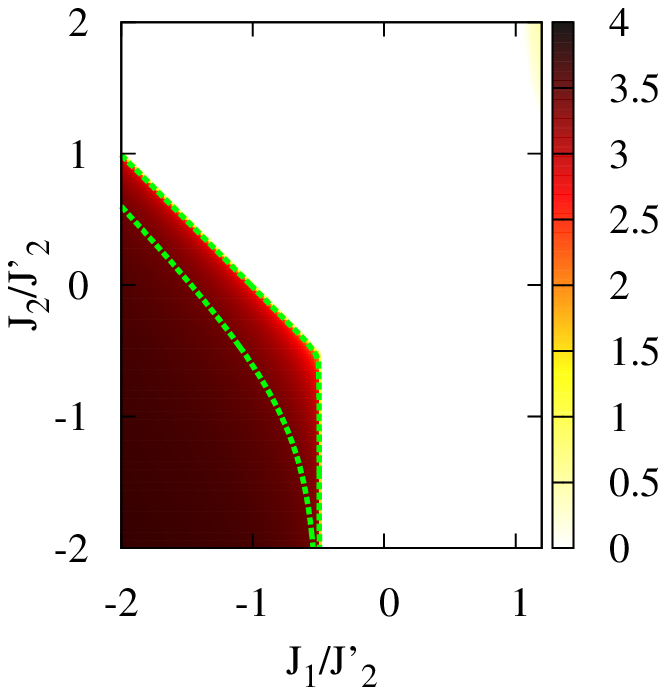}
\label{fig:8b}
}
\label{fig:8}
\caption[]{Contour maps of magnetization $m$ (in units of $\mu_B$) in spin-wave calculations for $J^\prime_1=J^\prime_2$ (in (a)) and $J^\prime_1=-J^\prime_2$ (in (b)). The dashed green lines are the contour lines of $m=2.0\mu_B$ per Fe (upper line) and $m=3.4\mu_B$ per Fe (lower line). The region in between defines the physical parameter regime.
}
\end{figure}

It is worth discussing which regime in the parameter space is most relevant to the experimentally observed block-spin state. To address this issue we may define a physical parameter regime in the AF1 phase by requiring the theoretically calculated magnetization, $m$, to be within the range of magnetic moments determined in experiments. Experimentally, it is found that the magnetic moments take values between $2\mu_B$ and $3.4\mu_B$ for various $A_y\mathrm{Fe}_{1.6}\mathrm{Se}_2$ ($A = \mathrm{K}, \mathrm{Tl}, \mathrm{Cs}, \mathrm{Rb}$) compounds.\cite{Yeetal} Theoretically, $m=S-\delta m$, where $\delta m = \frac{S}{4}\sum_{\alpha,\beta=1,2,3,4} \int_{\mathbf{k}\in \mathrm{MBZ}} |V_{\alpha\beta,\mathbf{k}}|^2$, is the correction to the magnetization due to quantum fluctuations. We show the contour maps of $m$ from spin-wave calculations for $J^{\prime}_1=J^{\prime}_2$ and $J^{\prime}_1=-J^{\prime}_2$ in Fig.~\ref{fig:8a} and Fig.~\ref{fig:8b}, respectively. In both cases, we find the physical parameter regime is only limited within a narrow region near the boundary of the AF1 state. Note that the LDA calculation also suggests the exchange coupling parameters are close to the phase boundary between AF1 and AF2 states.\cite{Cao_Dai_block_spin}

\begin{figure}[htbp]
\centering
\subfigure[]{
\includegraphics[scale=0.2,
bbllx=30pt,bblly=0pt,bburx=761pt,bbury=584pt
]{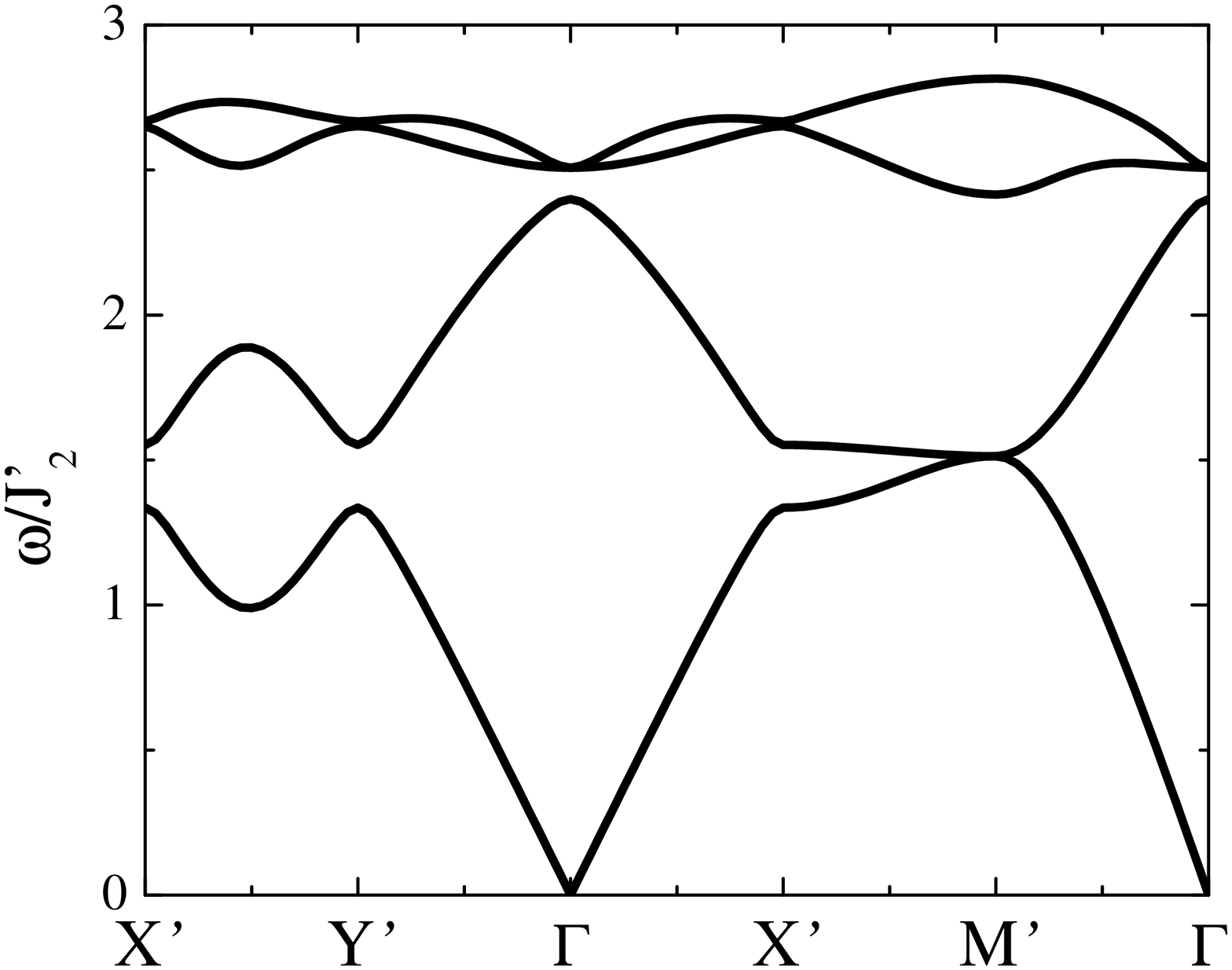}
\label{fig:4a}
}
\subfigure[]{
\includegraphics[scale=0.2,
bbllx=30pt,bblly=0pt,bburx=761pt,bbury=584pt
]{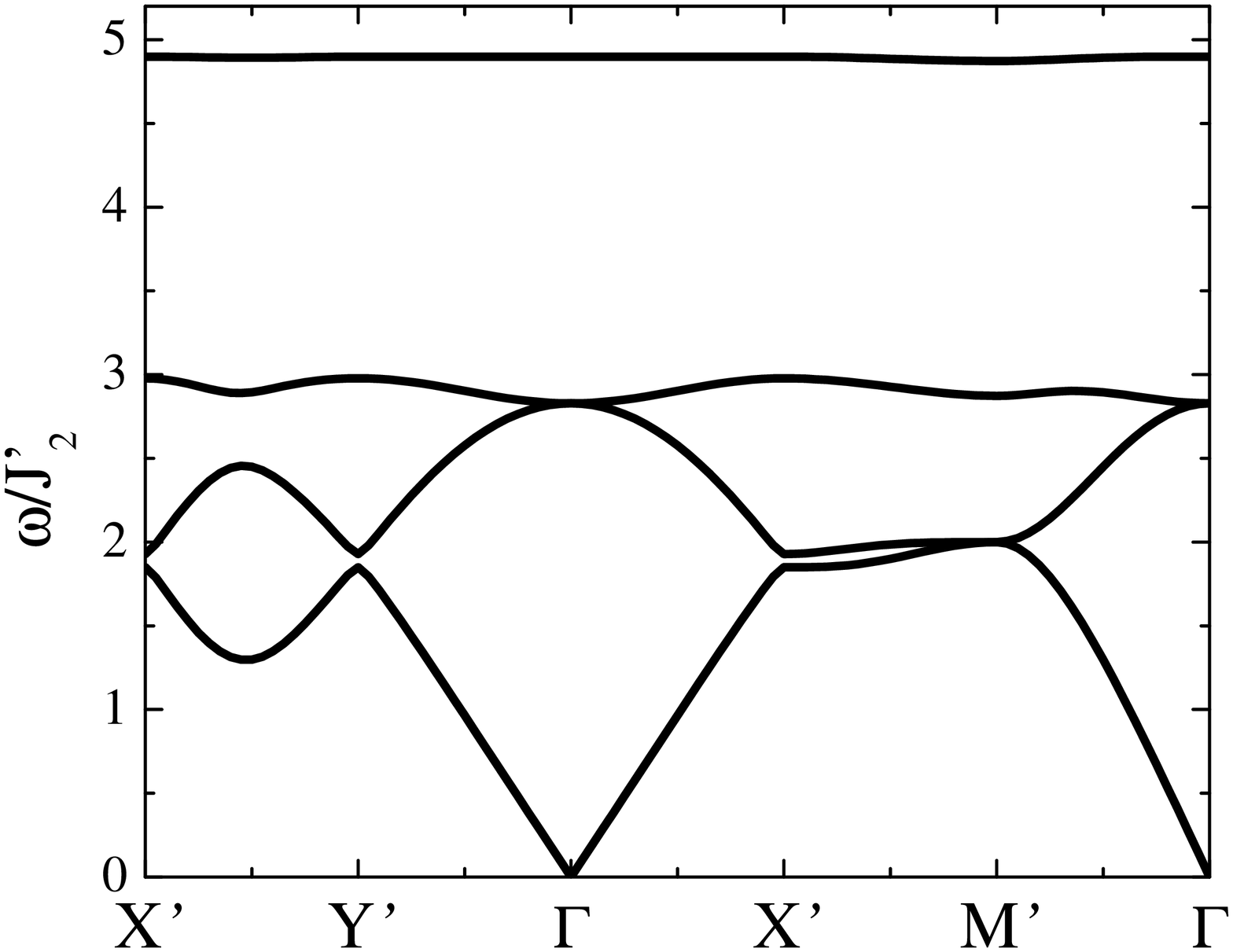}
\label{fig:4b}
}
\subfigure[]{
\includegraphics[scale=0.2,
bbllx=30pt,bblly=0pt,bburx=761pt,bbury=584pt
]{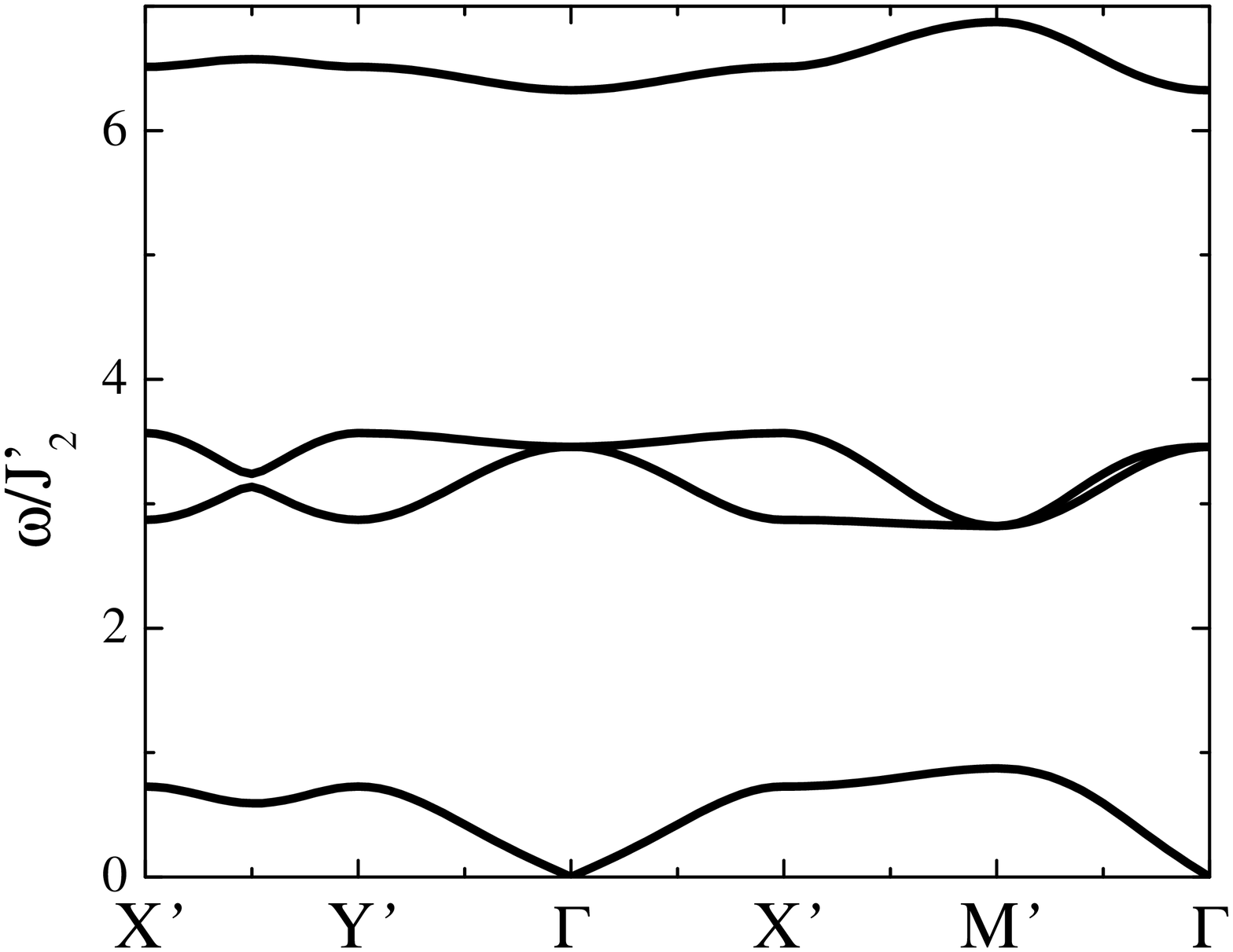}
\label{fig:4c}
}
\label{fig:4}
\caption[]{Spin-wave spectra along the high-symmetry directions of the MBZ1 of the block spin state for three representative sets of parameters: a) $J_1=0.1J'_2$, $J_2=0.05J'_2$, $J'_1=J'_2$; b) $J_1=-0.5J'_2$, $J_2=0.5J'_2$, $J'_1=J'_2$; c) $J_1=-1.5J'_2$, $J_2=0.2J'_2$, $J'_1=-J'_2$. In all three cases, $J'_2=1$, is the energy unit.
}
\end{figure}

To discuss the spin-wave spectra, we further follow the suggestion by LDA calculation~\cite{Cao_Dai_block_spin} to assume both $J_2$ and $J'_2$ to be antiferromagnetic, and show in Fig.~\ref{fig:4a} to Fig.~\ref{fig:4c} the calculated spin-wave
dispersions for three representative points in the phase diagram. The spin-wave dispersions are obtained from the eigenenergies in MBZ1. Due to the four Fe sites within a block, in addition to the gapless acoustic branch (Goldstone mode), there are also three gapped optical branches. Each of these four branches are also doubly degenerate.
Working out the eigenenergies at $\Gamma$ and M$^\prime$ points in MBZ1, we find that the degeneracy of the eigenenergies at these two points is helpful to determining the sign and relative strength of the exchange couplings. As shown in Fig.~\ref{fig:4a}, when $J_1$ and $J'_1$ are both positive, the top two optical branches at $\Gamma$ point are degenerate, and the lowest optical branch and the acoustic branch at M$^\prime$ point are degenerate. In this case, there is a finite gap between the top two optical branches and the rest of the spin-wave spectrum. If $J_1<0$ and $J_2\sim|J_1|$, as shown in Fig.~\ref{fig:4b}, the lowest two optical branches at $\Gamma$ point are degenerate, and the lowest optical branch and the acoustic branch at M$^\prime$ point are also degenerate. In this case, a finite gap separates the top optical branches from the rest of the spin-wave dispersion. On the other hand, when $J_1<0$ but $J_2\ll|J_1|$, the lowest two optical branches at both $\Gamma$ and M' points are degenerate. In this case, the acoustic branch is completely separated from the optical ones. Moreover, a finite gap also separates the top optical branches from the two lower optical branches, as shown in Fig.~\ref{fig:4c}.

The spin-wave spectrum for Rb$_y$Fe$_{1.6}$Se$_2$ has been recently measured through neutron scattering experiment.~\cite{PDai_spin_wave_block} In the experimental spin-wave dispersion the acoustic branch is well separated from the optical ones, is similar to the one shown in Fig.~\ref{fig:4c}. The dispersion is well fitted to the extended $J_1$-$J_2$ model in Eq.~\ref{eq:1} with $J_1<0$ and $J_2\ll |J_1|$. This confirms the direct relevance of our results to the experiments.

\section{$J_1-J_2$ model for $2\times2$ vacancy ordered $L_2$ lattice and $4\times2$ vacancy ordered $L_3$ lattice}
\begin{figure}[htbp]
\centering
\subfigure[]{
\includegraphics[scale=0.2,
bbllx=80pt,bblly=40pt,bburx=641pt,bbury=584pt
]{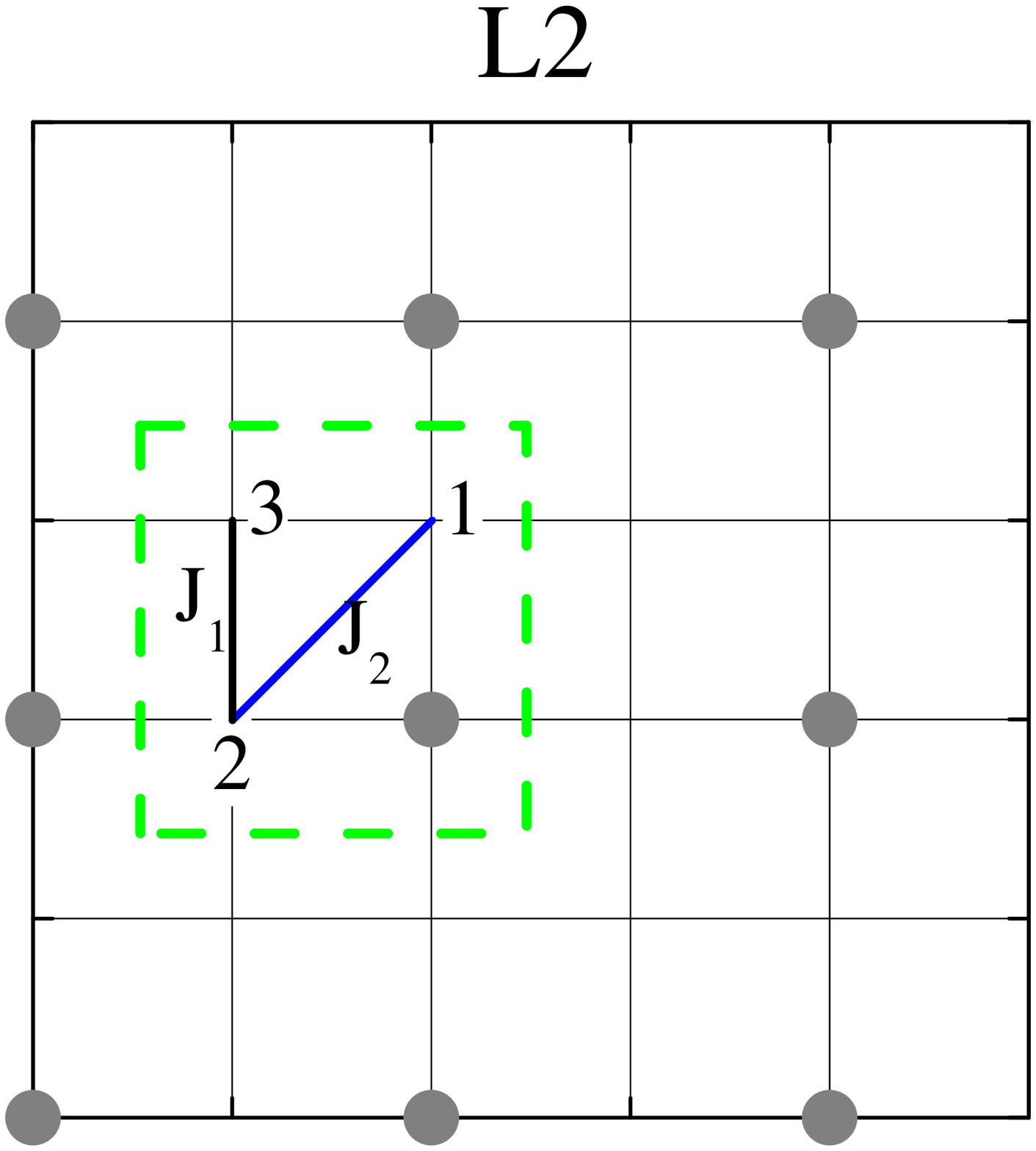}
\label{fig:5a}
}
\subfigure[]{
\includegraphics[scale=0.2,
bbllx=80pt,bblly=40pt,bburx=641pt,bbury=584pt
]{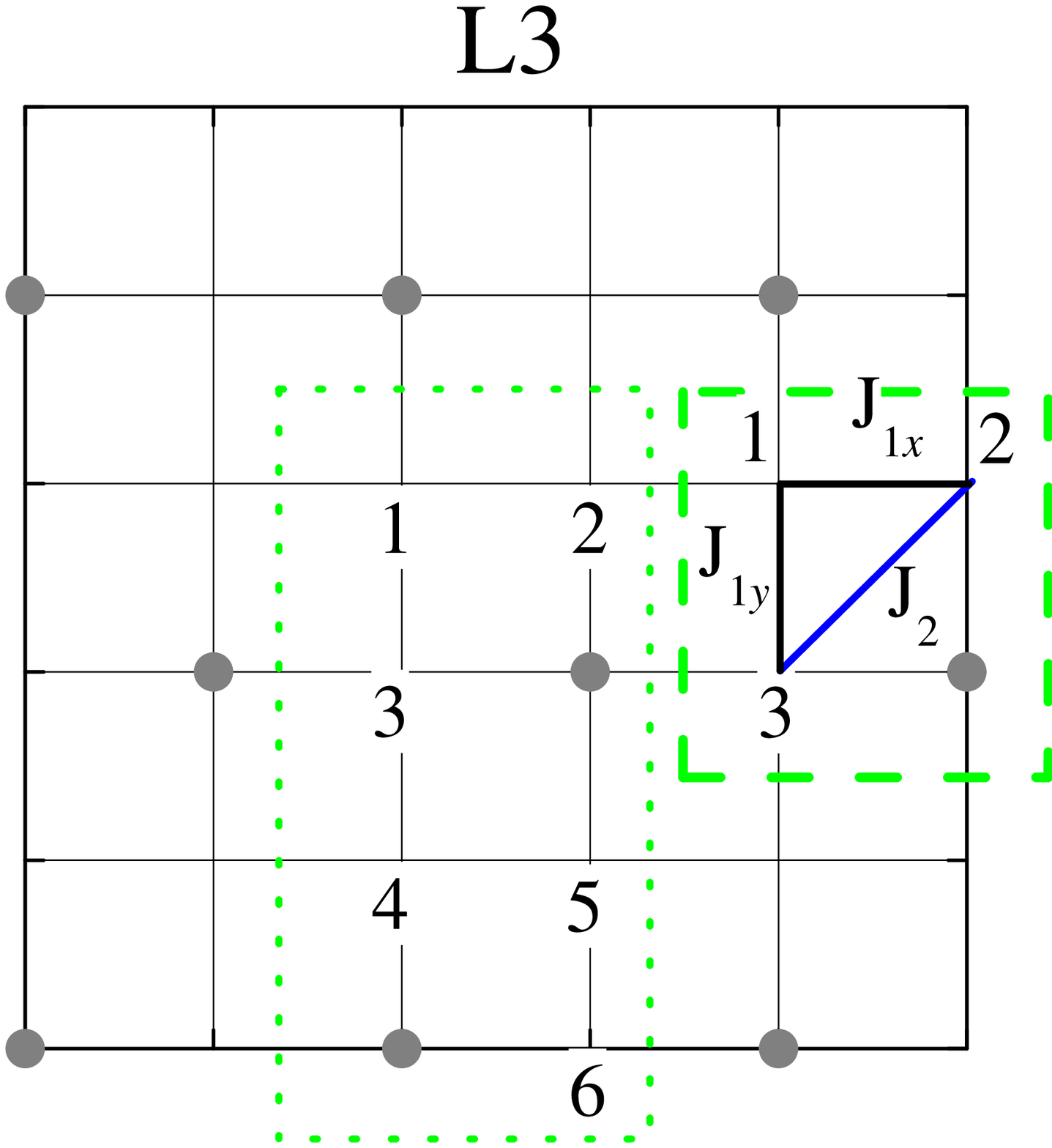}
\label{fig:5b}
}
\label{fig:5}
\caption[]{In panels (a) and (b) we respectively show $2\times2$ vacancy ordered $L_2$ lattice and $4\times 2$ vacancy ordered $L_3$ lattice. In panel (a) the magnetic unit cell for $L_2$ lattice is marked by dashed green line, and the magnetic unit cell consists of three $\mathrm{Fe}$ sites denoted as $1, 2, 3$. In panel (b) the three-Fe magnetic unit cell for $L_3$ lattice is shown using dashed green lines, and we also show a larger unit cell that consists of six $\mathrm{Fe}$ sites denoted as $1,2,...,6$.}
\end{figure}
In this section we discuss the magnetic phase diagrams for two different modulated lattices $L_2$ and $L_3$ which respectively possess $2\times2$ and $4\times2$ vacancy orders. These two vacancy orders have been claimed to be important for $\mathrm{K}_y\mathrm{Fe}_{1.5}\mathrm{Se}_2$. In Fig.~ \ref{fig:5a} and Fig.~\ref{fig:5b} we respectively show $L_2$ and $L_3$ lattices. We note that the $2 \times 2$ vacancy ordering preserves the $C_{4v}$ symmetry of the square lattice, and for this reason we analyze an isotropic $J_1-J_2$ model for $L_2$ lattice. However, $4\times2$ vacancy order breaks the $C_{4v}$ symmetry, and for this reason we analyze an anisotropic $J_{1x}-J_{1y}-J_2$ model for $L_3$ lattice.
The explicit form of the Hamiltonian is given by
\begin{eqnarray}
H_2 &=& J_{1x}\sum_{i} \mathbf{S}_{i}\cdot \mathbf{S}_{i+\hat{x}}+J_{1y}\sum_{i} \mathbf{S}_{i}\cdot \mathbf{S}_{i+\hat{y}}\nonumber \\
&+& J_2\sum_{i} \mathbf{S}_{i}\cdot \mathbf{S}_{i+\hat{x\pm y}},
\end{eqnarray}
where $J_{1x}=J_{1y}$ for $L_2$. For the description of different magnetic states on $L_2$ we use a $2\times2$ unit cell including three Fe sites, as shown in Fig.~ \ref{fig:5a}  with dashed green lines. To describe the magnetic states on $L_3$, we can use a similar three-Fe unit cell shown as shown in Fig.~\ref{fig:5b} with dashed green lines. Sometimes it is more convenient to understand the nature of the magnetic state on $L_3$ by using a larger $4\times2$ six-Fe unit cell as shown in Fig.~\ref{fig:5b} with dotted green line.

\subsection{Magnetic phases and phase diagram for $L_2$ lattice} 
\begin{figure}[htbp]
\centering
\subfigure[]{
\includegraphics[scale=0.2,
bbllx=80pt,bblly=10pt,bburx=641pt,bbury=564pt
]{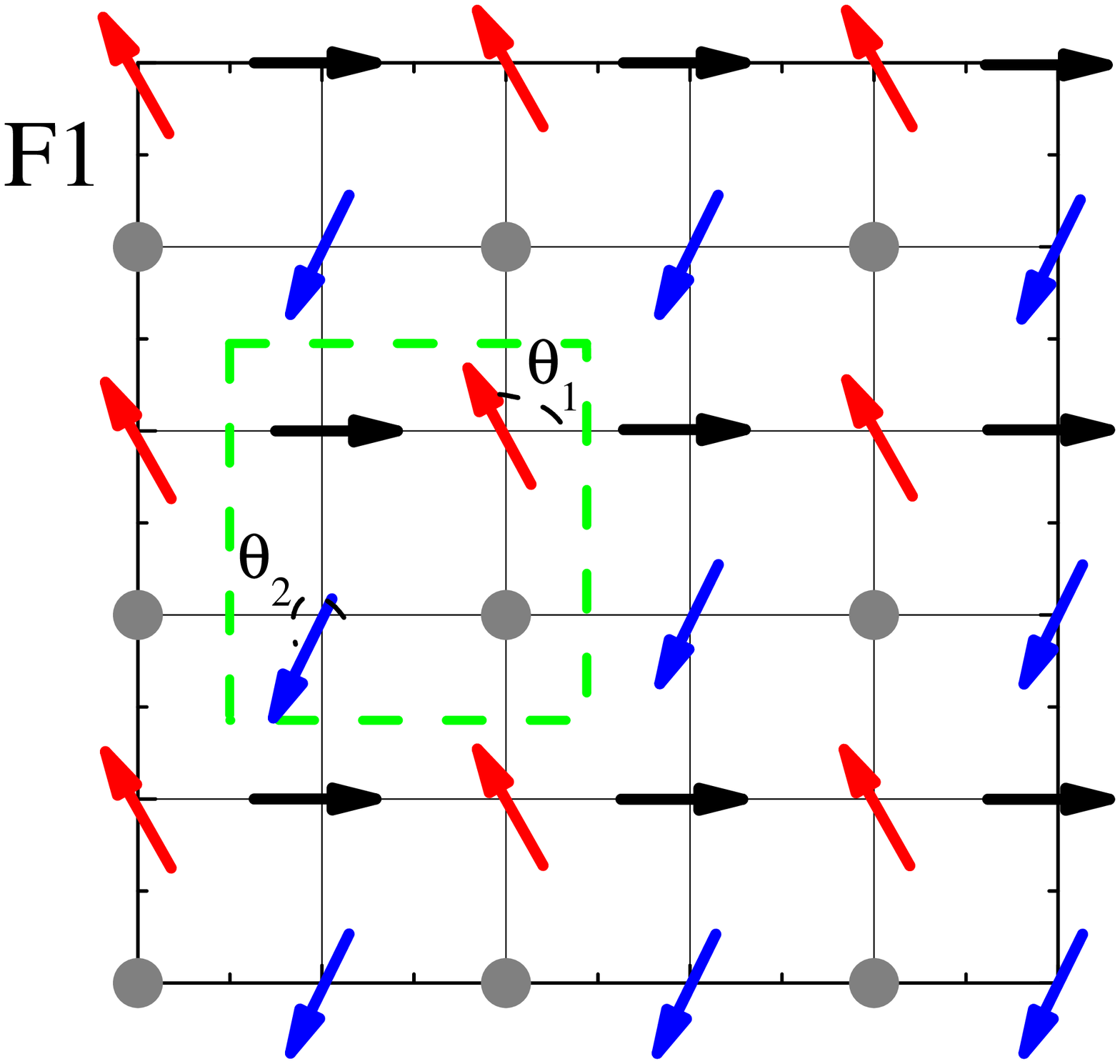}
\label{fig:6a}
}
\subfigure[]{
\includegraphics[scale=0.2,
bbllx=80pt,bblly=10pt,bburx=641pt,bbury=564pt
]{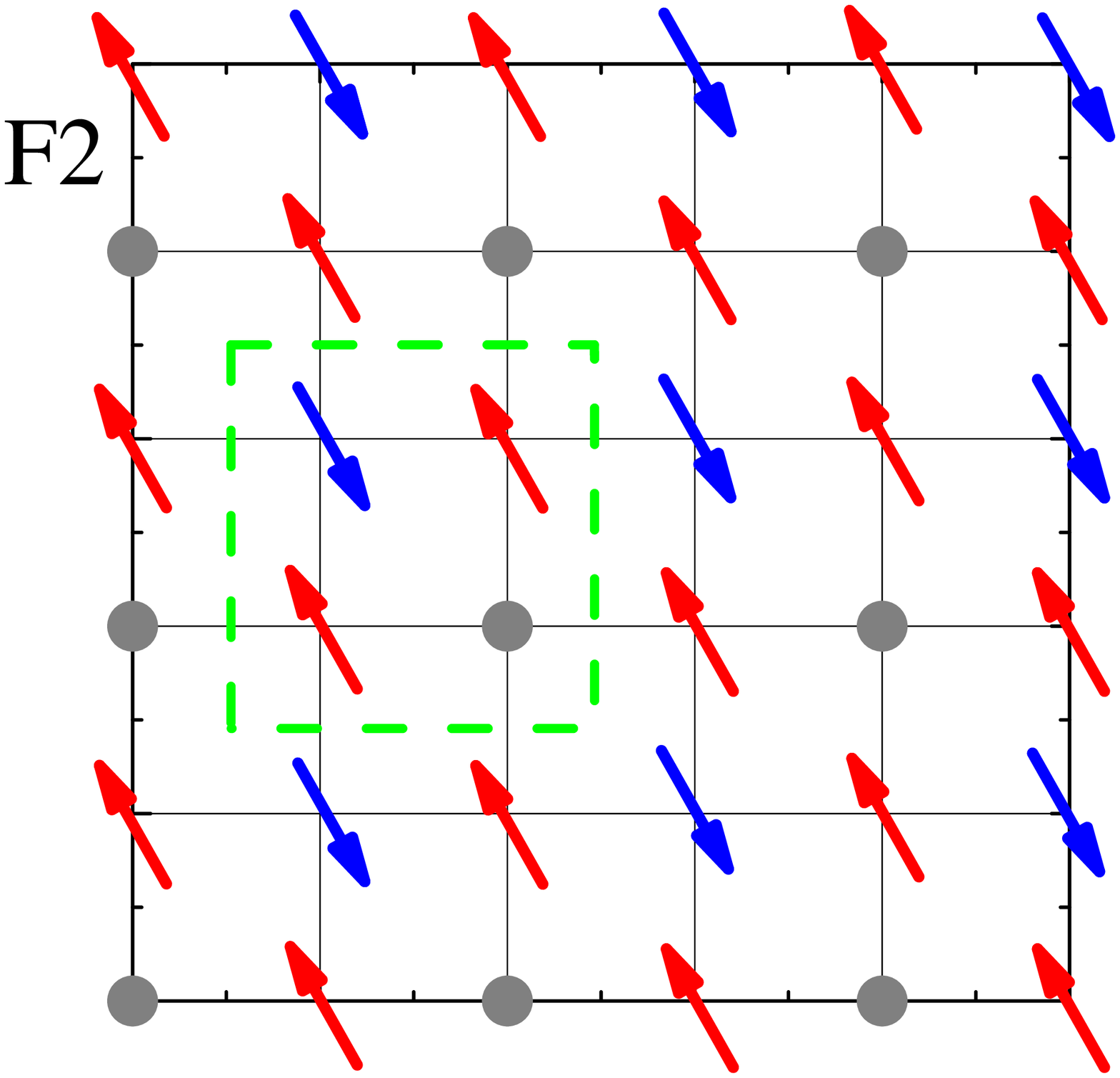}
\label{fig:6b}
}
\subfigure[]{
\includegraphics[scale=0.2,
bbllx=70pt,bblly=10pt,bburx=631pt,bbury=564pt
]{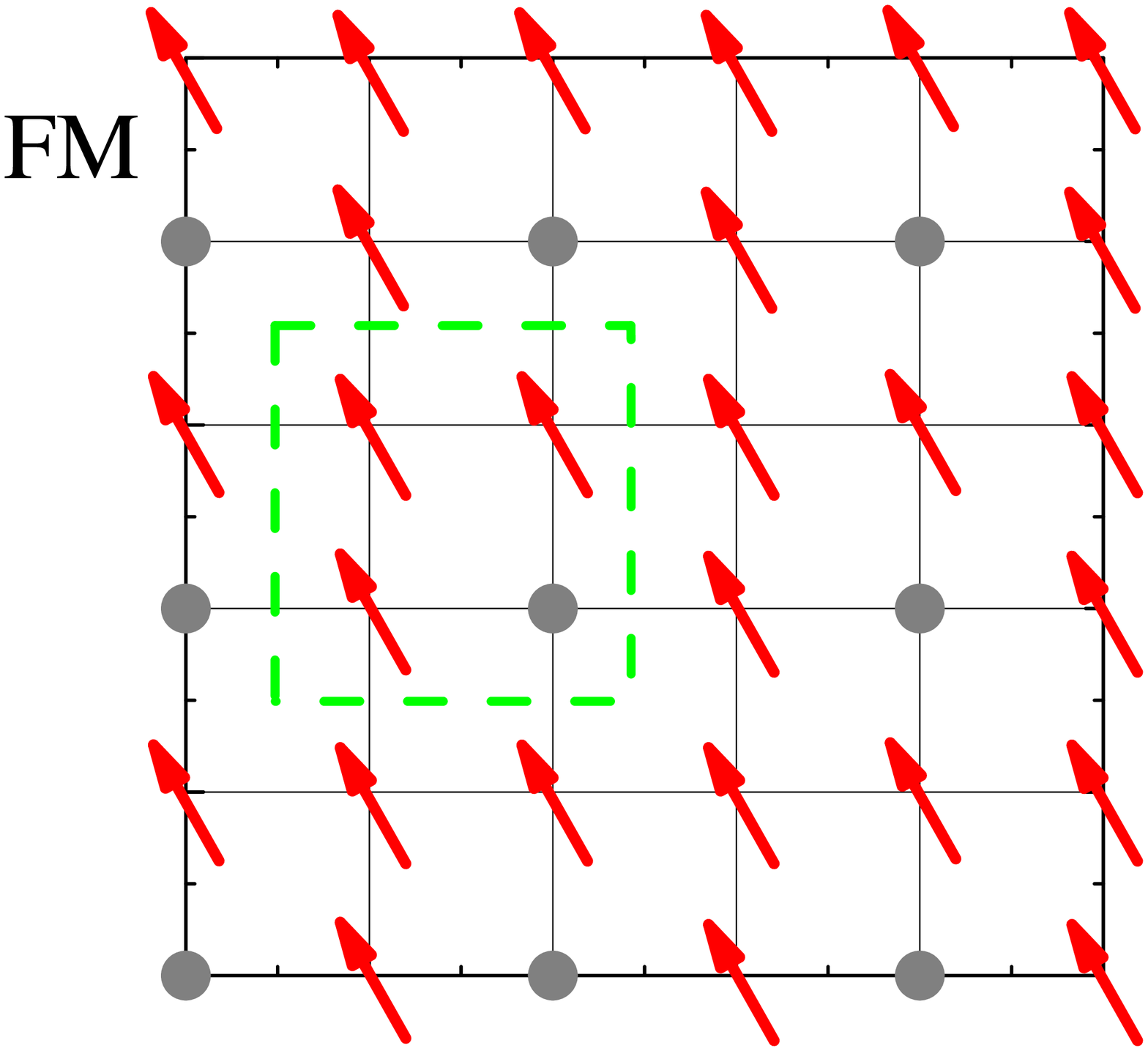}
\label{fig:6c}
}
\subfigure[]{
\includegraphics[scale=0.21,
bbllx=80pt,bblly=100pt,bburx=641pt,bbury=464pt
]{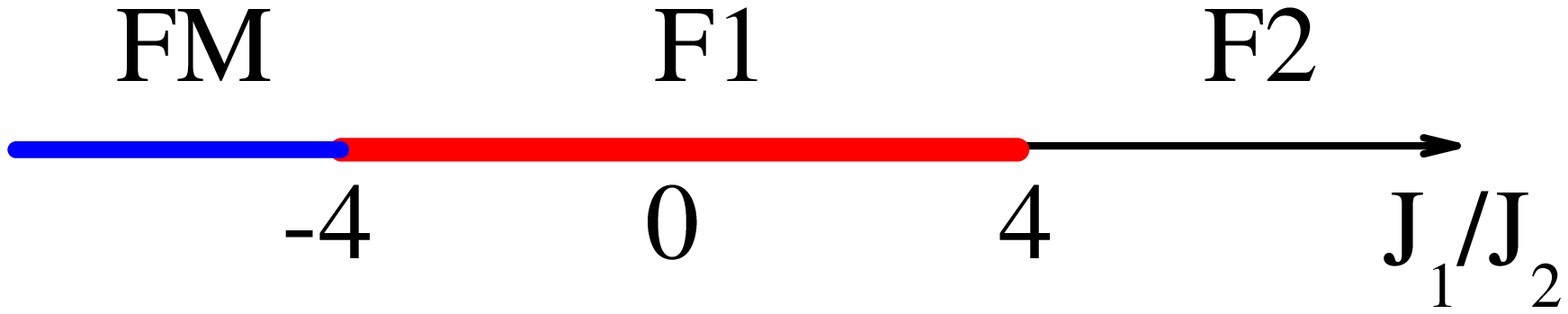}
\label{fig:6d}
}
\label{fig:6}
\caption[]{In panels (a), (b), (c) we respectively show three possible phases F1, F2, F3 found for the $L_2$ lattice. F1 and F2 are respectively non-collinear and collinear ferrimagnetic states, and F3 is the conventional ferromagnetic state. In panel (d) we show the relevant phase diagram as a function of the ratio $J_1/J_2$.
}
\end{figure}

We again employ a classical Monte Carlo technique to understand the phase diagram for $L_2$ lattice. There are three phases F1, F2, FM and these are respectively shown in Fig.~\ref{fig:6a}, Fig.~\ref{fig:6b}, and Fig.~\ref{fig:6c}. The phase diagram as a function of the ratio $J_1/J_2$ is shown in Fig.~\ref{fig:6d}. These three phases are commensurate in terms of three-Fe magnetic unit cell, and all three phases have ordering vector $\mathbf{Q}=(0,0)$. All three states can be described by a single mode spiral ansatz, as introduced for $L_1$. The values of $\mathbf{Q}$, the canting angles $\theta_{\alpha}$, and energy per unit site for all three phases are listed in TABLE.~\ref{tab:2}. For general $J_1/J_2$ ratio, the F1 state is a non-collinear ferrimagnetic state. But it turns out to be a non-collinear antiferromagnetic state if $J_1=2J_2$. F2 state has the conventional Neel arrangement, but the presence of vacancy gives rise to nonzero magnetic moment per unit cell. For this reason, F2 is a collinear ferrimagnetic phase. Finally FM is the conventional, collinear ferromagnetic state. The phase boundaries are found by using the energy values listed TABLE.~\ref{tab:2}.

\begin{center}
\begin{table}[htdp]
\begin{tabular}{|c|c|c|c|} 
\hline
Phase & $\mathbf{Q}$-vector & $\theta_{\alpha}$ & Energy per site \\
\hline\cline{1-4}
FM & $\mathbf{Q}=(0,0)$ & $\theta_{1,2,3}=0$ & $4(J_1+J_2)/3$\\
\hline
F1 & $\mathbf{Q}=(0,0)$ & $\theta_{1}=\cos^{-1}(-J_1/4J_2)$ & $-(8J_2^2+J_1^2)/6$\\
& & $\theta_{2}=-\theta_{1}$, $\theta_{3}=0$& \\
\hline
F2 & $\mathbf{Q}=(0,0)$ & $\theta_{1,2}=\pi$, $\theta_{3}=0$ & $4(J_2-J_1)/3$\\
\hline
\end{tabular}
\caption{The ordering vector $\mathbf{Q}$ in the MBZ for three Fe site unit cell, the canting angles and the ground-state energies per Fe site for the magnetic states shown in Fig.~\ref{fig:6a} through Fig.~\ref{fig:6c}.}\label{tab:2}
\end{table}
\end{center}

\subsection{Magnetic phases and phase diagram for $L_3$ lattice}
\begin{figure}[htbp]
\centering
\subfigure[]{
\includegraphics[scale=0.17,
bbllx=80pt,bblly=10pt,bburx=641pt,bbury=564pt
]{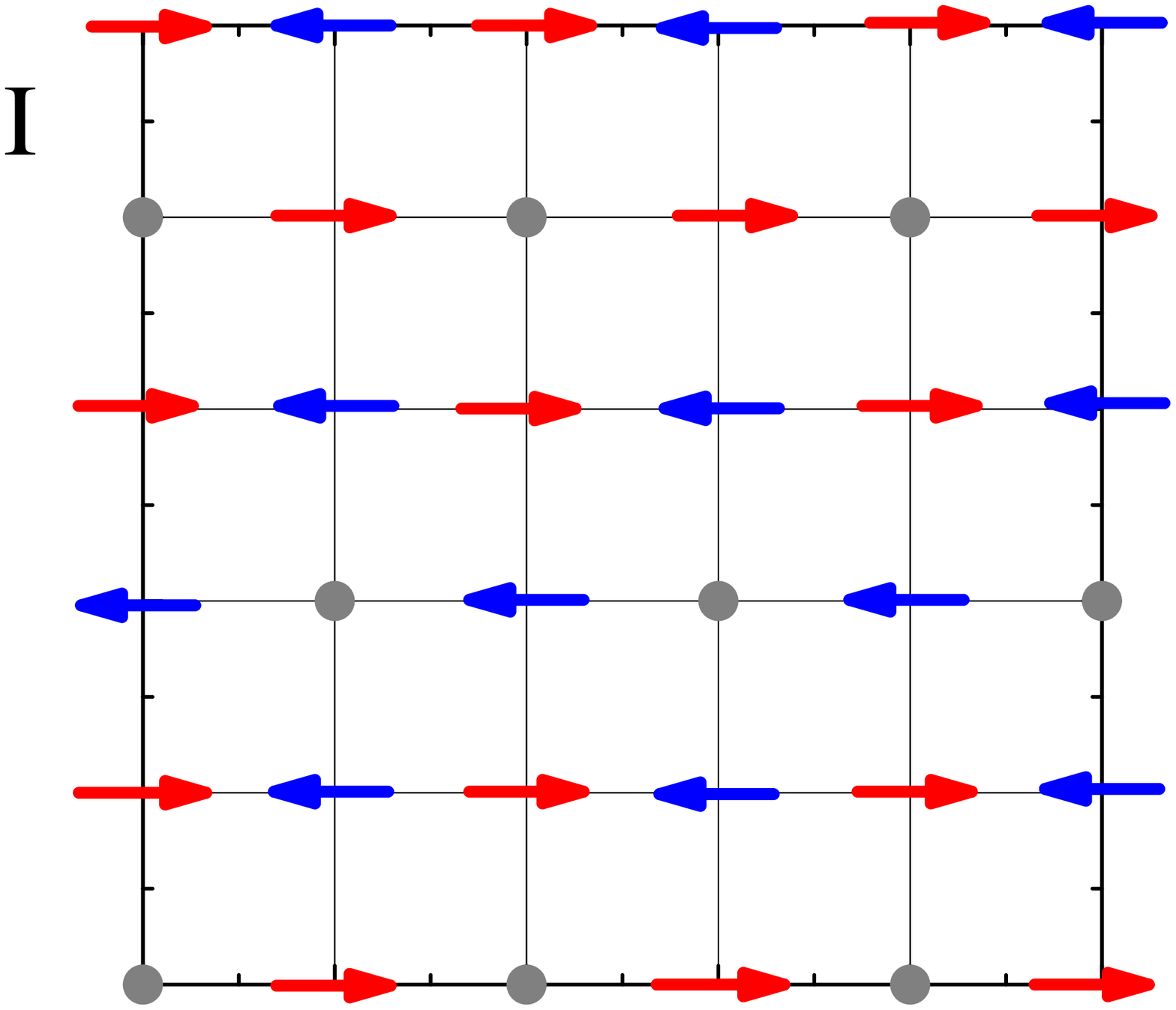}
\label{fig:7a}
}
\subfigure[]{
\includegraphics[scale=0.17,
bbllx=80pt,bblly=10pt,bburx=641pt,bbury=564pt
]{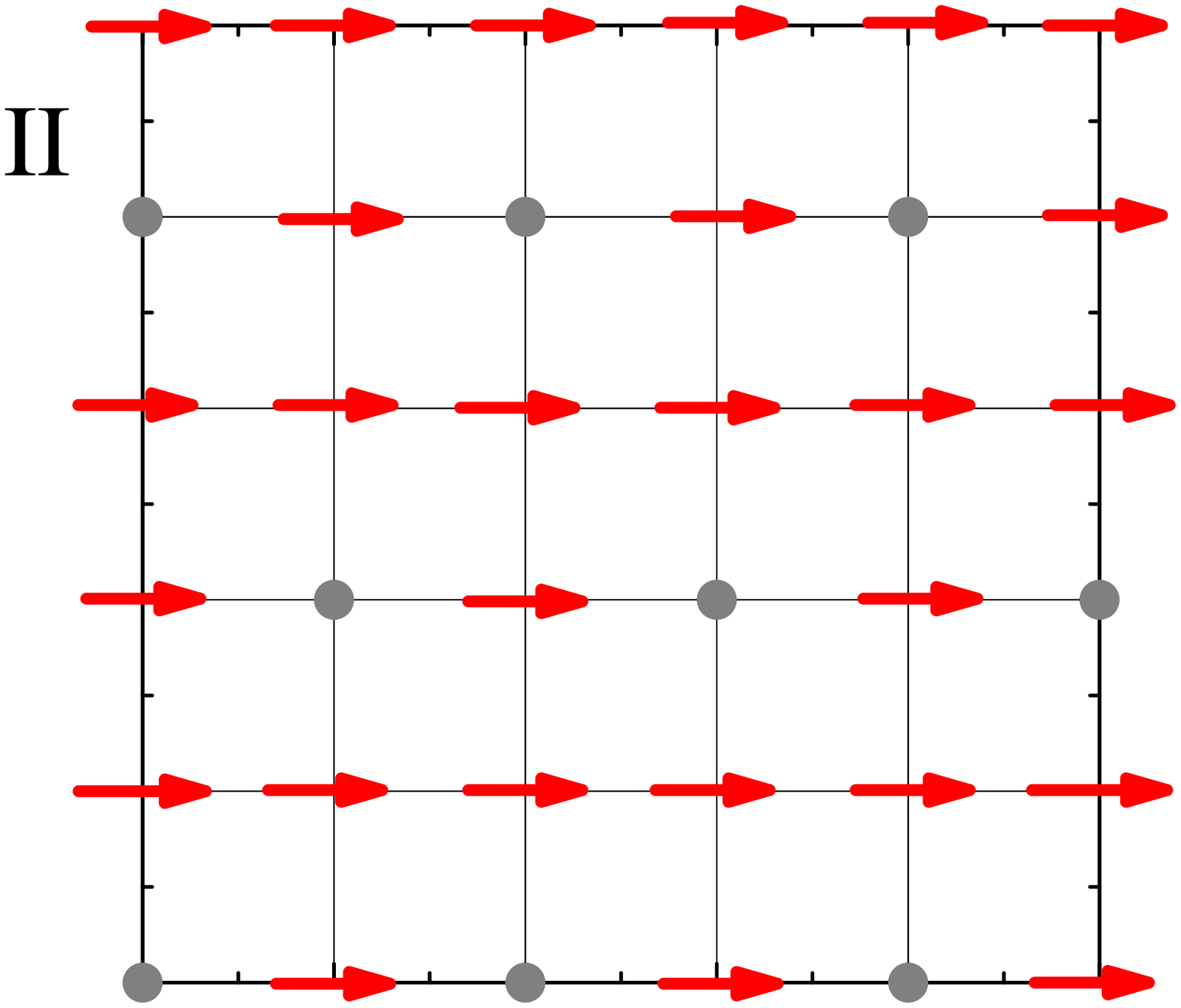}
\label{fig:7b}
}
\subfigure[]{
\includegraphics[scale=0.17,
bbllx=80pt,bblly=10pt,bburx=641pt,bbury=564pt
]{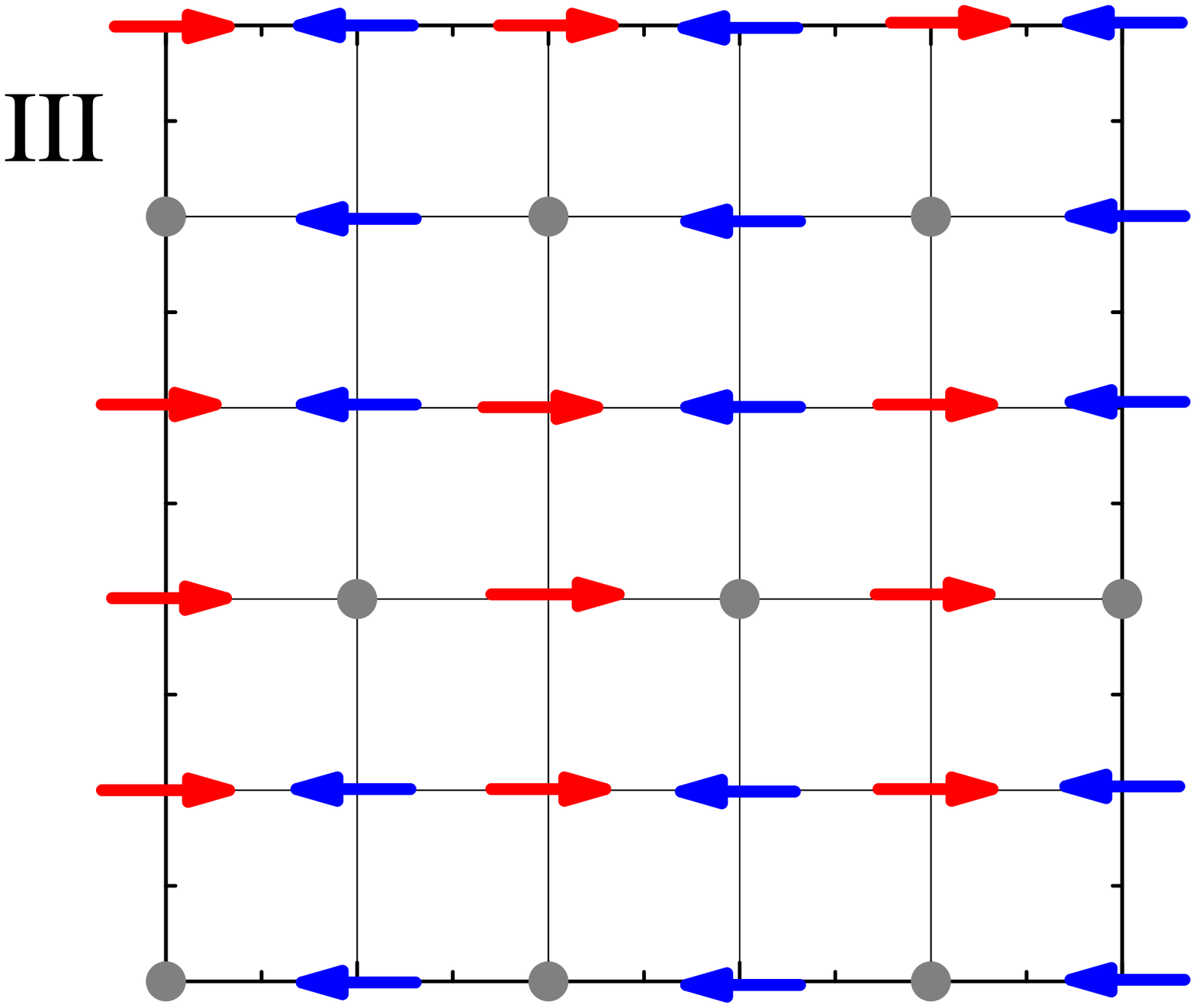}
\label{fig:7c}
}
\subfigure[]{
\includegraphics[scale=0.17,
bbllx=80pt,bblly=10pt,bburx=641pt,bbury=564pt
]{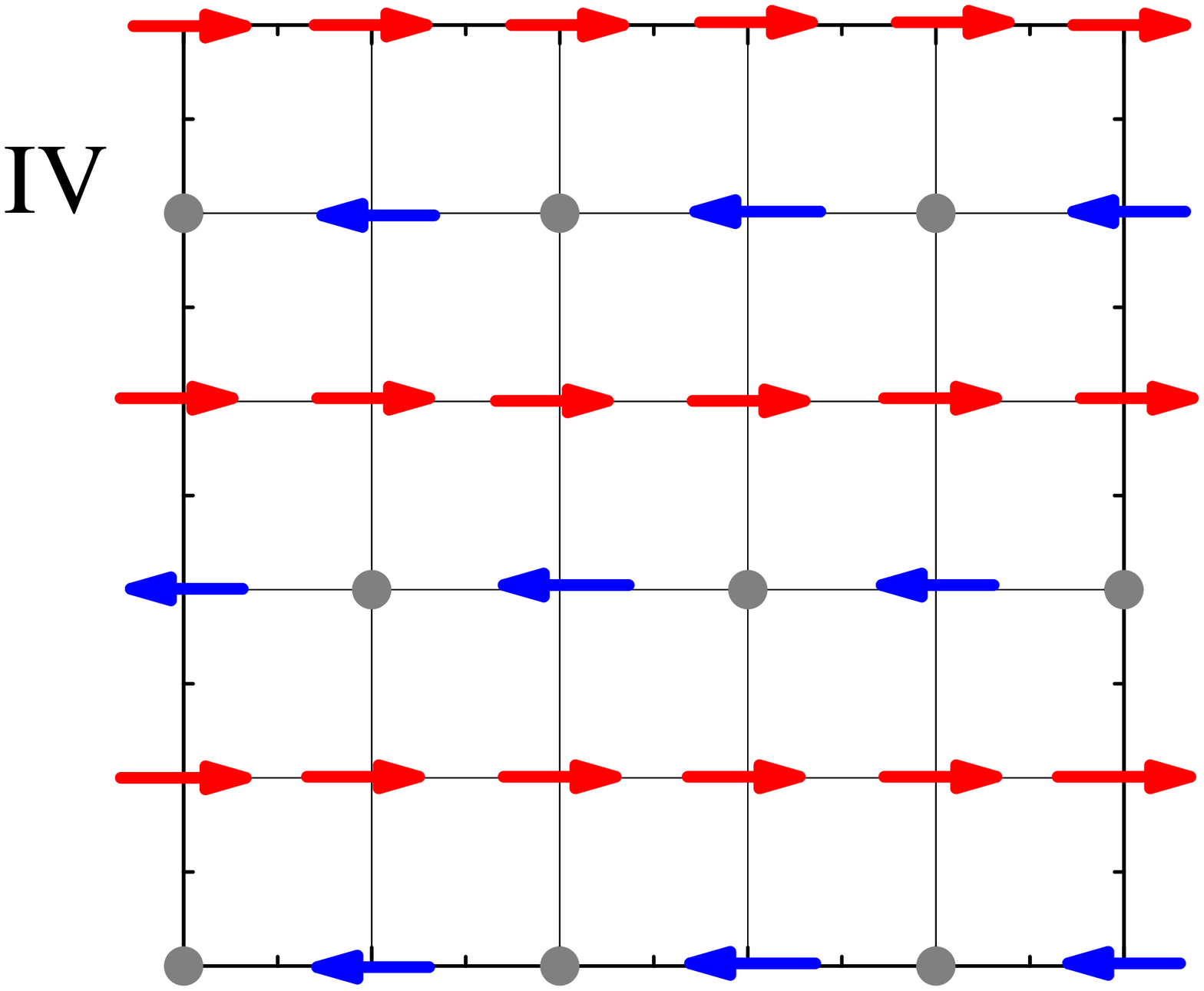}
\label{fig:7d}
}
\subfigure[]{
\includegraphics[scale=0.17,
bbllx=80pt,bblly=10pt,bburx=641pt,bbury=564pt
]{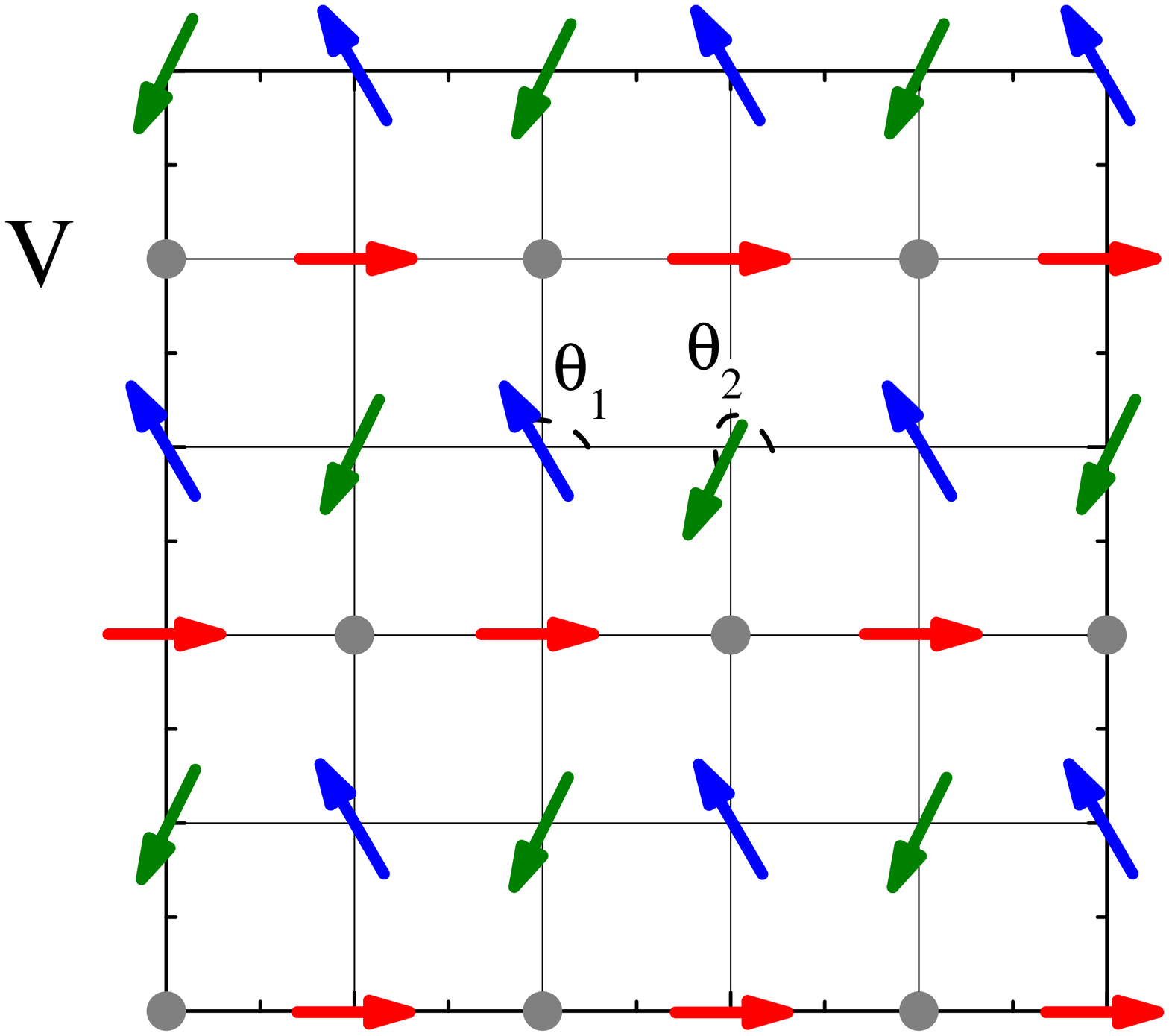}
\label{fig:7e}
}
\subfigure[]{
\includegraphics[scale=0.17,
bbllx=80pt,bblly=10pt,bburx=641pt,bbury=564pt
]{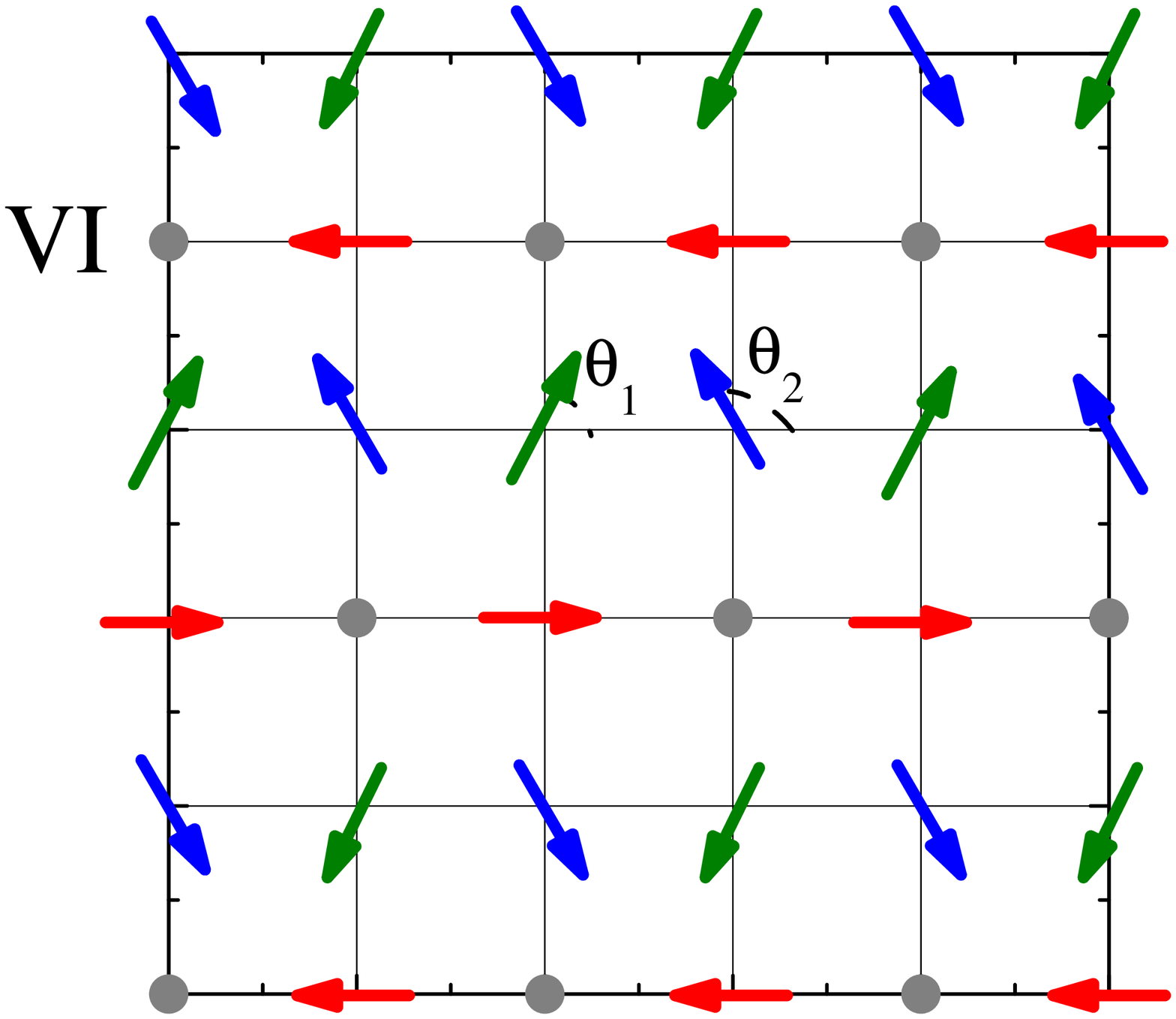}
\label{fig:7f}
}
\subfigure[]{
\includegraphics[scale=0.17,
bbllx=80pt,bblly=10pt,bburx=641pt,bbury=564pt
]{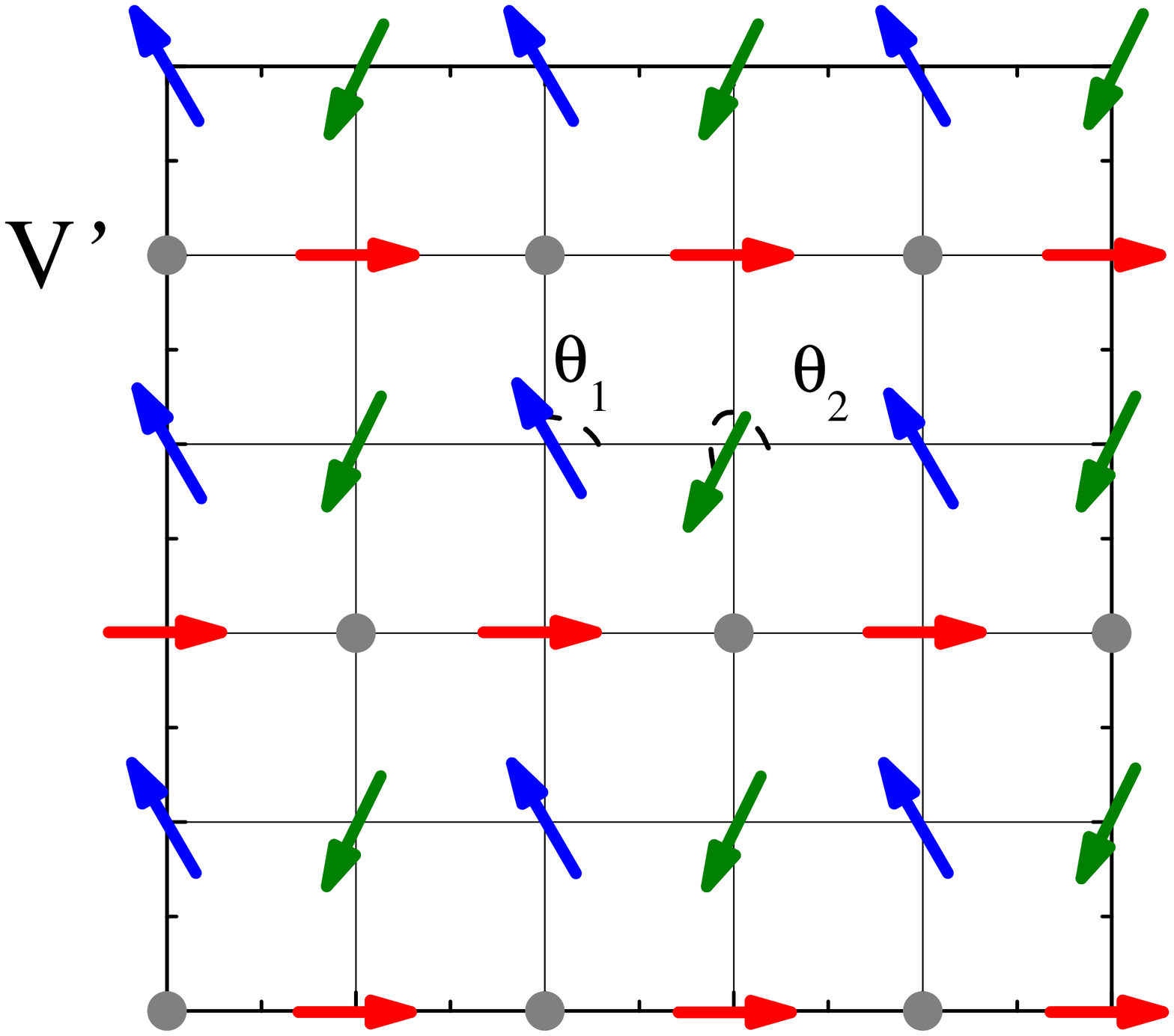}
\label{fig:7g}
}
\subfigure[]{
\includegraphics[scale=0.17,
bbllx=80pt,bblly=10pt,bburx=641pt,bbury=564pt
]{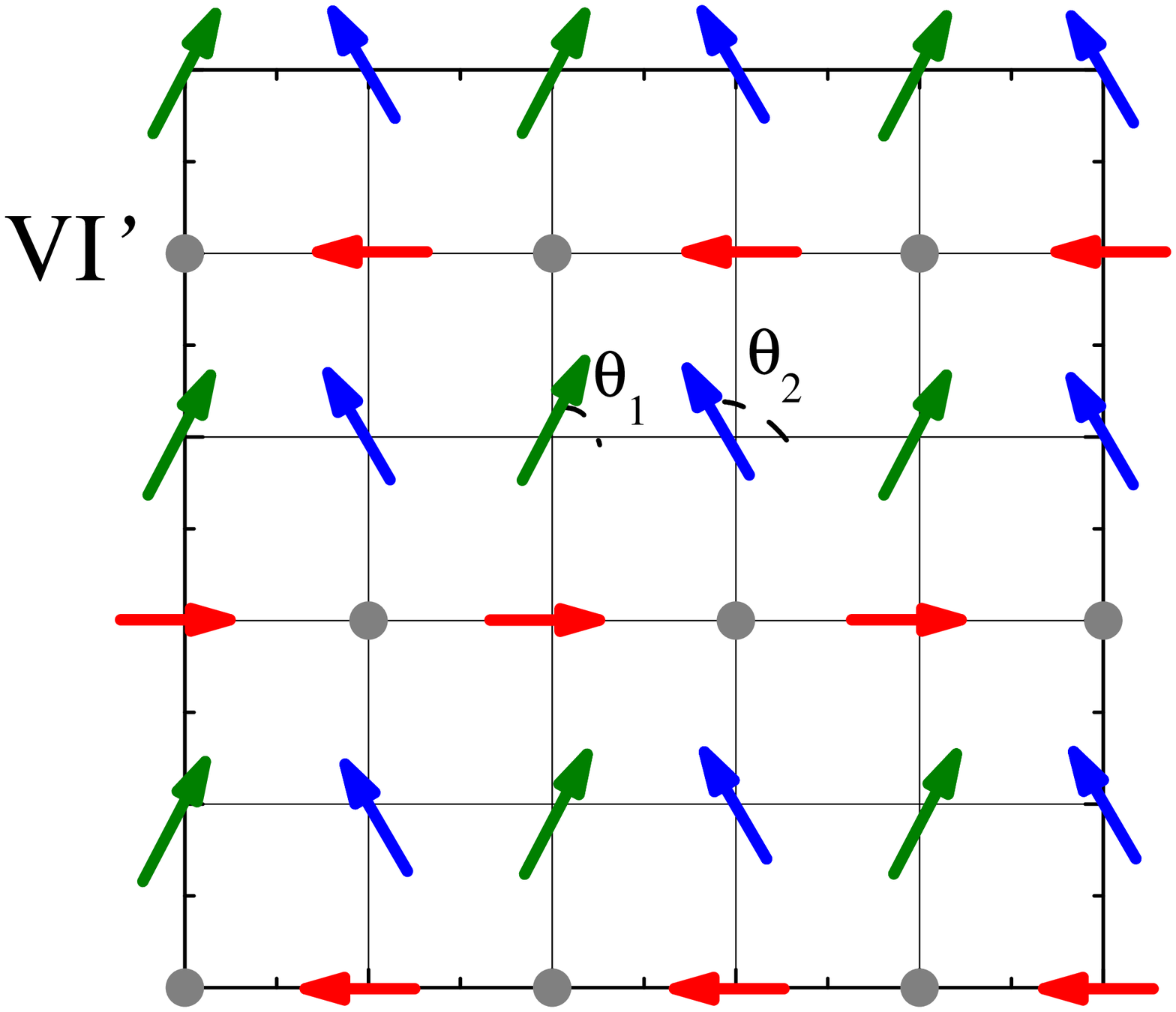}
\label{fig:7h}
}
\subfigure[]{
\includegraphics[scale=0.22,
bbllx=80pt,bblly=10pt,bburx=641pt,bbury=564pt
]{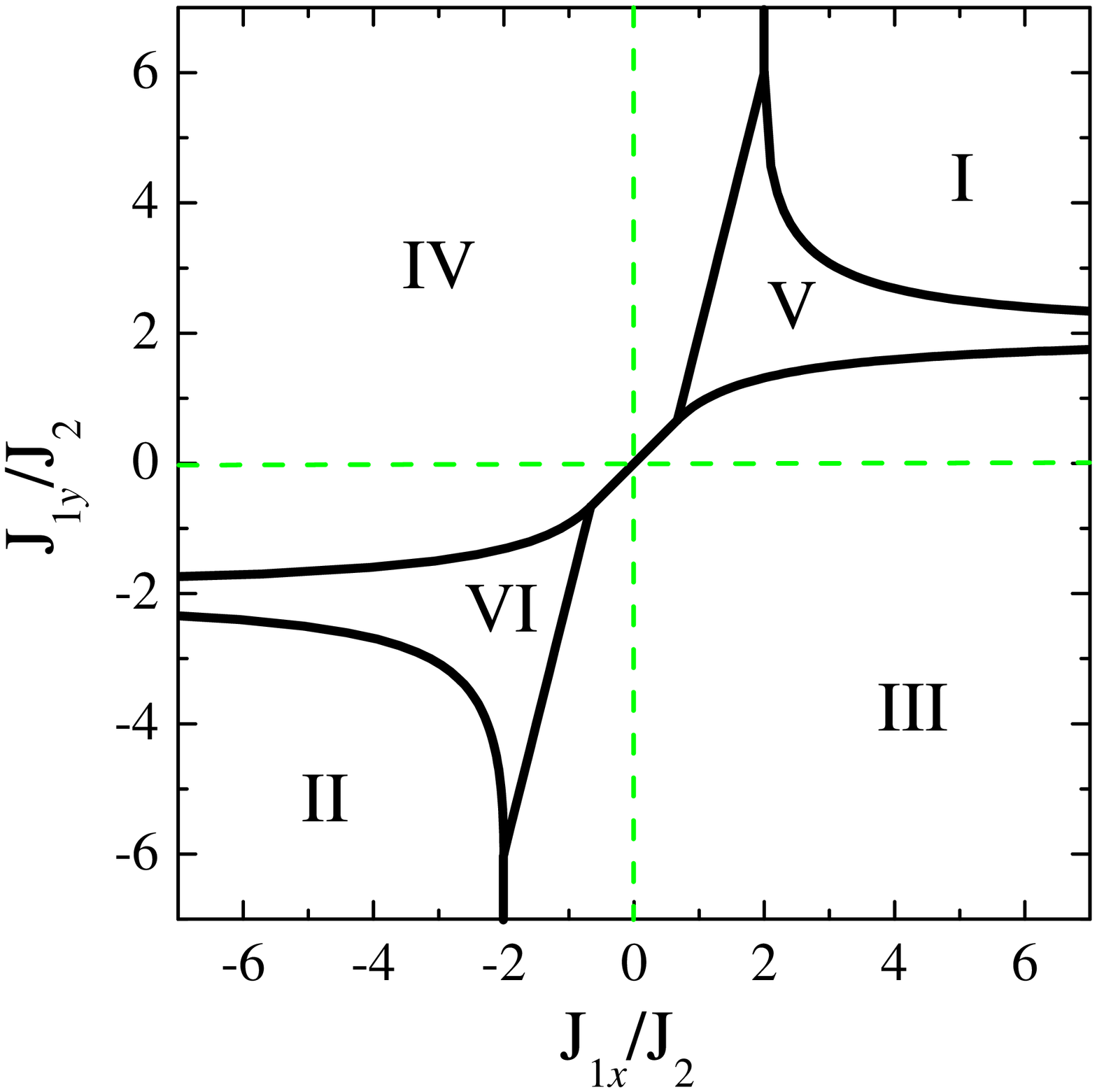}
\label{fig:7i}
\label{fig:7}
}
\caption[]{In panels (a) through (h) we show the spin arrangements for possible magnetic states I through VI' realized for $L_3$ lattice. States I, II, III, and IV are collinear and the states V, V', VI, VI'are non-collinear. The states IV and V are ferrimagnetic. The states I, III, and VI are antiferromagnetic, and the state II is ferromagnetic. The pairs of states (V, V') and (VI, VI') are degenerate, and coexist in the phase diagram displayed in panel (i).
}
\end{figure}

The possible magnetic states for the $L_3$ lattice are again found from a classical Monte Carlo calculation. There are six phases denoted as I through VI, and the spin arrangements in these phases are shown in Fig.~\ref{fig:7a} through Fig.~\ref{fig:7h}. In the MBZ corresponding to three-Fe unit cell the spin states I through VI have either $(0,0)$ or $(0,\pi)$ wavevector. Interestingly, we find in phases V and VI the pairs of states (V,V') and (VI, VI') are energetically degenerate, and coexist in the phase diagram. In Fig.~\ref{fig:7i} we show the phase diagram in the $J_{1x}/J_2-J_{1y}/J_2$ plane. The wavevectors, canting angles and energy per site for states I to VI are displayed in TABLE.~\ref{tab:3}.

The magnetic states I, II, III and IV have collinear spin arrangements,
whereas V, V',VI, VI' have non-collinear spin arrangements. The state I is an antiferromagnet with the spins having conventional Neel arrangement. The state II is the conventional ferromagnetic state. The state III has conventional $(\pi, 0)$ antiferromagnetic arrangement, and even in the presence of the vacancy order, this state remains antiferromagnetic. To see this one needs to consider the large $4\times 2$ unit cell with six Fe sites. The state IV has conventional $(0, \pi)$ antiferromagnetic arrangement. But, the vacancy order leads to nonzero magnetic moment for $4 \times 2$ unit cell, and this state becomes ferrimagnetic. Interestingly, state IV and state III are energetically degenerate when $J_{1x}=J_{1y}$, though the $L3$ lattice breaks the $C_4$ symmetry. This degeneracy is not lifted even by introducing a finite $J_3$ coupling between the 3rd nearest neighbor spins. The state V is a non-collinear ferrimagnetic state, and state VI is a non-collinear antiferromagnetic state. Both of them can be described by the single mode spiral ansatz and have commensurate ordering wavevectors in the three-Fe MBZ. But states V' and VI' cannot be described by the single mode spiral ansatz in the three-Fe MBZ, though they are degenerate with V and VI, respectively. By comparing Fig.~\ref{fig:7e} with Fig.~\ref{fig:7g} we see that state V' can be obtained from V by shifting all spins in the $4n+1$ rows to the left by one lattice spacing. Similarly, state VI' can be obtained from VI by first flipping all spins in the $4n+1$ rows and then shifting them to the left by one lattice spacing. Note that both state V' and state VI' can be described by the single mode spiral ansatz and have commensurate ordering wavevector $(0,0)$ in the BZ corresponding to the $4 \times 2$ six-Fe unit cell. Both state V' and state VI' are non-collinear ferrimagnetic.

\begin{center}
\begin{table}[htdp]
\begin{tabular}{|c|c|p{3.4cm}|p{2.8cm}|} 
\hline
Phase & $\mathbf{Q}$-vector & $\quad\quad\quad\quad\theta_{\alpha}$ & $\quad$Energy per site \\
\hline\cline{1-4}
I & $\mathbf{Q}=(0,\pi)$ & $\theta_{1}=\pi$, $\theta_{2,3}=0$ & $2(2J_2-J_{1x}-J_{1y})/3$\\
\hline
II & $\mathbf{Q}=(0,0)$ & $\theta_{1,2,3}=0$ & $2(J_{1x}+J_{1y}+2J_2)/3$\\
\hline
III & $\mathbf{Q}=(0,\pi)$ & $\theta_{2}=\pi$, $\theta_{1,3}=0$ & $2(J_{1y}-J_{1x}-2J_2)/3$\\
\hline
IV & $\mathbf{Q}=(0,0)$ & $\theta_{1,2}=\pi$, $\theta_{3}=0$ & $2(J_{1x}-J_{1y}-2J_2)/3$\\
\hline
V & $\mathbf{Q}=(0,0)$ & $\theta_{1}=\cos^{-1}\left(-\frac{2J_2+J_{1y}}{4J_{1x}}\right)$ & $-[8J_{1x}^2+(2J_2+J_{1y})^2]/12J_{1x}$\\
& & $\theta_{2}=-\theta_{1}$, $\theta_{3}=0$ & \\
\hline
VI & $\mathbf{Q}=(0,\pi)$ & $\theta_{1}=\cos^{-1}\left(-\frac{2J_2-J_{1y}}{4J_{1x}}\right)$& $[8J_{1x}^2+(2J_2-J_{1y})^2]/12J_{1x}$\\
& & $\theta_{2}=\pi-\theta_{1}$, $\theta_{3}=0$ &\\
\hline
\end{tabular}
\caption{The ordering wavevector $\mathbf{Q}$ in three-Fe MBZ, the canting angles, and the ground-state energies per Fe site for the magnetic states shown in Fig.~\ref{fig:7a} through Fig.~\ref{fig:7h}.}\label{tab:3}
\end{table}
\end{center}

\section{Summary and Conclusion}
To summarize, we have studied the magnetic phase diagram of an extended $J_1-J_2$ model on several modulated square lattices. For a $\frac{1}{5}$-depleted square lattice with $\sqrt{5} \times \sqrt{5}$ vacancy order we have shown that a block-spin antiferromagnetic state, of the type observed in
$\mathrm{K}_{y}\mathrm{Fe}_{1.6}\mathrm{Se}_2$, arises over a significant region of the phase diagram.
This region comprises three parts: a) all the intra-block exchange couplings ($J_1$,$J_2$) and inter-block couplings ($J_1^{\prime}$,$J_2^{\prime}$) are antiferromagnetic; b) $J_2$, $J_1^{\prime}$,$J_2^{\prime}$
are antiferromagnetic, while $J_1$ is ferromagnetic; c) $J_1^{\prime}$ and $J_2^{\prime}$
are antiferromagnetic, while $J_1$ and $J_2$ are ferromagnetic.

We have also calculated the spin-wave spectrum and the renormalized magnetic moment in the block-spin state, from which we locate the most experimentally relevant parameter regime in the phase diagram. By studying the spin-wave spectrum in this experimentally relevant parameter regime, it is
suggested that measurements of spin gaps and degeneracy along the
high symmetry directions of the magnetic Brillouin zone will provide
valuable information regarding which part of the parameter space the
exchange couplings belong to.

Finally, we have studied the magnetic phase diagram of a $J_1-J_2$ model on $\frac{1}{4}$-depleted square lattices with $2 \times 2 $  or $4 \times 2$ vacancy orders. These phase diagrams are likely to be relevant for $\mathrm{K}_y\mathrm{Fe}_{1.5}\mathrm{Se}_2$.

This work was supported by
NSF Grant No. DMR-1006985 and the Robert A. Welch Foundation
Grant No. C-1411.
During the final stage of writing up this paper, related works on the magnetism
of $\mathrm{K}_{y}\mathrm{Fe}_{1.6}\mathrm{Se}_2$ have appeared
\cite{Zhou_Yao_Lee, Fang_etal, Lu_Dai}.

\end{document}